\documentclass[10pt]{article}

\usepackage{amsmath, amssymb}

\usepackage{graphicx}

\usepackage{cite}
\usepackage{color} 

\usepackage[labelfont=bf,labelsep=period,justification=raggedright]{caption}

\makeatletter
\renewcommand{\@biblabel}[1]{\quad#1.}
\makeatother

\date{}

\begin{document}

% Title must be 150 characters or less
\begin{flushleft}
{\Large
\textbf{Can retinal ganglion cell dipoles seed iso-orientation domains in the visual cortex?}
}
% Insert Author names, affiliations and corresponding author email.
\\
Manuel Schottdorf$^{1-5}$, 
Stephen J. Eglen$^6$,
Fred Wolf$^{1-4}$, 
Wolfgang Keil$^{1-4,7,\ast}$
\\
\bf{1} Max Planck Institute for Dynamics and Self-Organization, 37077 G\"ottingen, Germany
\\
\bf{2} Institute for Nonlinear Dynamics, University of G\"ottingen, 37077 G\"ottingen, Germany
\\
\bf{3} Bernstein Center for Computational Neuroscience, 37077 G\"ottingen, Germany
\\
\bf{4} Bernstein Focus for Neurotechnology, 37077 G\"ottingen, Germany
\\
\bf{5} Institute for Theoretical Physics, University of W\"urzburg, 97074, W\"urzburg, Germany
\\
\bf{6} Department of Applied Mathematics and Theoretical Physics, Cambridge Computational Biology Institute, University of Cambridge, Cambridge, UK
\\
\bf{7} Center for Studies in Physics and Biology, The Rockefeller University, New York, NY, 10065, USA
\\
$\ast$ E-mail: wkeil@mail.rockefeller.edu
\end{flushleft}

% Please keep the abstract between 250 and 300 words
\section*{Abstract}
It has been argued that the emergence of roughly periodic orientation preference maps (OPMs) in the primary visual cortex (V1) of carnivores and primates can be explained by a so-called statistical connectivity model. This model assumes that input to V1 neurons is dominated by feed-forward projections originating from a small set of retinal ganglion cells (RGCs). The typical spacing between adjacent cortical orientation columns preferring the same orientation then arises via Moir\'{e}-Interference between hexagonal ON/OFF RGC mosaics. While this Moir\'{e}-Interference critically depends on long-range hexagonal order within the RGC mosaics, a recent statistical analysis of RGC receptive field positions found no evidence for such long-range positional order. 
Hexagonal order may be only one of several ways to obtain spatially repetitive OPMs in the statistical connectivity model. Here, we investigate a more general requirement on the spatial structure of RGC mosaics that can seed the emergence of spatially repetitive cortical OPMs, namely that angular correlations between so-called RGC dipoles exhibit a spatial structure similar to that of OPM autocorrelation functions. Both in cat beta cell mosaics as well as primate parasol receptive field mosaics we find that RGC dipole angles are spatially uncorrelated. To help assess the level of these correlations, we introduce a novel point process that generates mosaics with realistic nearest neighbor statistics and a tunable degree of spatial correlations of dipole angles. Using this process, we show that given the size of available data sets, the presence of even weak angular correlations in the data is very unlikely. We conclude that the layout of ON/OFF ganglion cell mosaics lacks the spatial structure necessary to seed iso-orientation domains in the primary visual cortex.
\section*{Introduction}
\begin{figure}[!ht]
\begin{center}
\includegraphics[width=10.0cm]{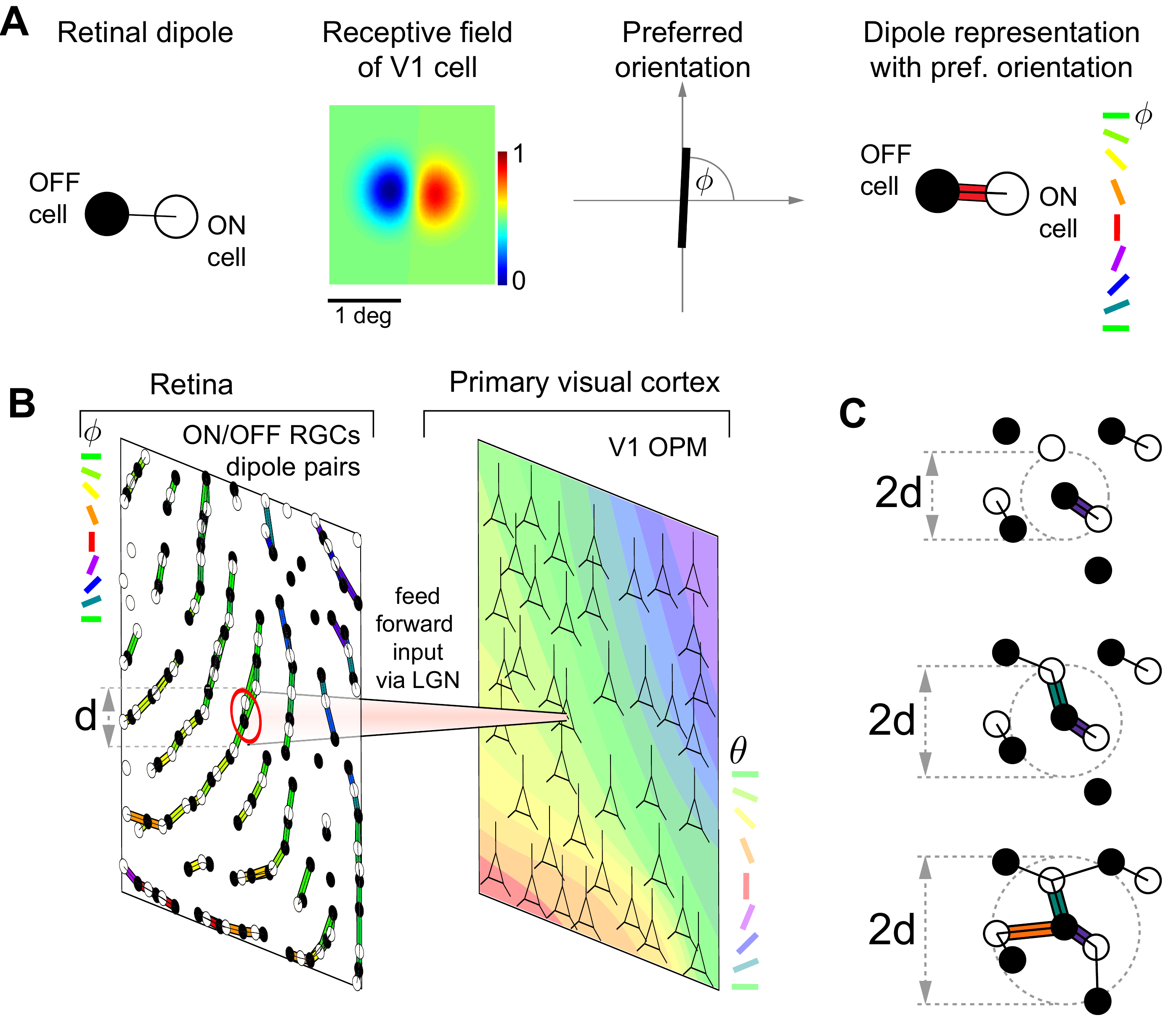}
\end{center}
\caption{\textbf{RGC dipoles and the statistical wiring model according to \cite{Paik2011}}
\textbf{A} Most left: A dipole of an ON center (empty circle) and OFF center (filled circle) retinal ganglion cell (RGC). The black line connecting the two cells indicates that the two cells form a dipole. A V1 cell with input dominated by this dipole has a receptive field with side-by-side subregions of opposite sign (middle left) and is consequently tuned to a specific orientation $\phi$ (middle right). We represent the preferred orientation of the V1 cell by the \textit{color} of the bar connecting the two RGCs (most right). Note that the preferred orientation of the V1 is orthogonal to the bar connecting the two RGCs.
\textbf{B} The statistical connectivity model for orientation preference maps. The receptive field midpoints of ON/OFF center RGCs are arranged in semiregular mosaics.The input to a cortical cell is dominated by a single pair of ON/OFF dipole and the cortical units have oriented receptive fields. If RGC dipole orientations are locally correlated, orientation preference $\theta$ within layer 4 of V1 is predicted to vary smoothly resulting in a smooth and continuous map of orientation preferences.
\textbf{C} Parametrized definition of RGC dipoles.  ON/OFF pairs with distance smaller than a parameter \textit{d} are considered dipoles (black lines). For the centered OFF cell, preferred orientations of dipoles are indicated. With increasing \textit{d}, the number of dipoles increases and one RGC can form multiple dipoles.
\label{fig1} }
\end{figure}
Many neurons in the primary visual cortex (V1) respond preferentially to edge-like stimuli of a particular orientation \cite{Hubel1962}. In carnivores and primates, orientation preference exhibits a columnar arrangement such that neurons positioned on top of each other from the white matter to the pia typically prefer similar orientations.
Tangential to the visual cortical layers, orientation preference changes smoothly and progressively \cite{Hubel1962} except at the centers of so-called pinwheels where neurons exhibiting the whole range of orientation preferences are located in close vicinity \cite{Hubel1962, Ohki2006}. The progression of orientation preferences across the visual cortical surface (Orientation preference map, OPM) appears as organized by a semiregularly spaced system of pinwheels and adjacent columns preferring the same orientation over roughly a millimeter distance  \cite{Blasdel1992a, Chapman1996, Crair1997, Obermayer1997, Bosking1997, Shmuel2000, Kaschube2009, Kaschube2010, Keil2012}.
\newline
Most models for the emergence of OPMs during postnatal development assume that their layout is determined by intracortical mechanisms (e.g. \cite{Durbin1990, Obermayer1992, Swindale1996, Wolf1998,Wolf2005, Keil2011}). However, several recent studies advance the notion that the structure of OPMs may result from a statistical wiring of feed-forward inputs from the mosaic of ON/OFF retinal ganglion cells (RGCs) to V1 \cite{Ringach2004a, Ringach2007, Paik2011, Paik2012}  (Fig. \ref{fig1}A), an idea pioneered by Soodak \cite{Soodak1986, Soodak1987}. ON/OFF ganglion cells are arranged in semiregular mosaics across the retina and project to the lateral geniculate nucleus (LGN) of the thalamus. Thalamic receptive fields resemble RGC receptive fields in shape, size, and spatial distribution \cite{Cleland1971, Cleland1985}. The retinotopic map \cite{Daniel1961a,Tusa1978, Tootell1982,Djavadian1983} allows neighboring retinal/thalamic ON and OFF center cells to project to neighboring neurons in the primary visual cortex. Most nearest neighbor RGCs are ON/OFF pairs \cite{Wassle1981a}. According to the statistical connectivity model, a V1 neuron predominantly samples feed-forward inputs from geniculate projections in its immediate vicinity \cite{Alonso2001}. 
If so, it is likely to receive input from a single pair of ON/OFF RGCs, a so-called \textit{dipole} \cite{Paik2011} (Fig. \ref{fig1}A, left). The neuron's receptive field would then be dominated by one ON and one OFF subregion (Fig. \ref{fig1}A, middle left) and its response orientation-tuned  \cite{Ringach2004a,Ringach2007, Paik2011}. In this picture, the preferred orientation is determined by the orientation of the RGC dipole (Fig. \ref{fig1}A, middle right, right). Consequently, a key prediction of the statistical connectivity model is that orientation preference across the surface of the primary visual cortex should mirror the spatial distribution of the ON/OFF dipole angles in the RGC mosaics \cite{Soodak1986, Soodak1987, Ringach2004a, Paik2011} (Fig. \ref{fig1}B). \newline
Paik and Ringach \cite{Paik2011,Paik2012} showed how this model can explain the experimentally observed roughly periodic structure of visual cortical OPMs. In their model, ON and OFF cell mosaics are assumed to form two independent noisy hexagonal lattices. Superposing these two lattices leads to a hexagonal pattern of dipole orientations via Moir\'{e} interference \cite{Amidror2009}. The statistical connectivity hypothesis then implies that this hexagonal pattern is mapped onto the cortex creating a roughly hexagonal OPM \cite{Paik2011}.\newline
Hore et al. recently cast substantial doubts on the presence of hexagonal order in experimentally measured RGC receptive field mosaics \cite{Hore2012}. They found that the positions of receptive fields of ON/OFF RGCs in monkey retina are inconsistent with long-range hexagonal order and are much better described by a so-called pairwise interacting point processes (PIPP) in a parameter regime where long-range positional correlations are absent \cite{Eglen2005, Hore2012, Zhan2000}. Such PIPP mosaics lack the long-range order necessary to create a Moir\'{e}-Interference pattern and hence OPMs predicted by the statistical connectivity model with PIPP mosaics do not exhibit the experimentally observed spatially repetitive arrangement of orientation columns\cite{Ringach2004a, Ringach2007, Hore2012}.\\
Moir\'{e}-Interference of hexagonal RGC lattices constitutes one particular way of creating an ordered arrangement of regularly spaced orientation columns in the statistical connectivity model. Other spatial arrangements of RGCs are conceivable that might lead to spatially repetitive cortical orientation maps resembling the ones experimentally measured.  Therefore, the lack of hexagonal structure in RGC mosaics found by Hore et al. does not \textit{per se} dismiss the statistical connectivity hypothesis. 
Here, we investigate a fundamental requirement on the spatial structure of RGC mosaics to seed the emergence of spatially repetitive cortical OPMs: a spatial correlation of RGC dipole angles across the retina. RGC dipole angle correlations are predicted to exhibit a spatial structure similar to that of OPM autocorrelation functions. This means that RGC dipole angles have to be locally positively correlated and anti-correlated on intermediate scales. The precise values for both of these scales depend on the column spacing of the OPM as well as the cortical magnification factor. We first systematically analyze two previously published cat beta cell mosaics \cite{Wassle1981a, Zhan2000} as well as one primate parasol receptive field mosaic \cite{Gauthier2009} with respect to their dipole angle correlation functions. In both species, we are unable to detect any statistically significant positive or negative correlation. Since all three mosaics analyzed each contain only around 100 cell positions (or RFs center positions), the absence of detectable correlations might be a consequence of the small size of the data sets. To address this question, we introduce a novel point process that generates mosaics with a tunable degree of spatial correlations of dipole angles. The spatial structure of these angular correlations is designed such as to match the autocorrelation functions of experimentally measured OPMs. At the same time, the local spatial statistical properties of the resulting model RGC mosaics agree well with the experiment. On the one hand, the mosaics generated by this process demonstrate that hexagonally organized RGCs are indeed not necessary to obtain realistic OPMs within the statistical connectivity model. On the other hand, the mosaics generate by the point process can be used as reference mosaics to access the amount of data needed to detect the presence or absence of correlations. Finally, by statistical comparison of mPIPP model mosaics and data, we show that, given the size of our data sets, the presence of even weak angular correlations in the data can be ruled out. We conclude, that the layout of ON/OFF ganglion cell mosaics apparently lack a fundamental feature necessary to explain the emergence of spatially repetitive orientation preference maps in V1 within the dipole approximation of the statistical connectivity model. Our results suggest that the ordered arrangement of orientation columns is unlikely to originate from the spatial layout of RGCs and, hence, likely set by intracortical mechanisms.\\
%LOOSE End
%When RGCs are modeled by noisy hexagonal lattices \cite{Paik2011}, the dipole pair orientations in the resulting Moir\'{e} interference pattern exhibit similarly shaped spatial correlations on short scales \cite{Paik2011,Paik2012b}. In fact, such correlations are a necessary prerequisite for the statistical connectivity model to reproduce OPMs with typical spacing between adjacent column preferring the same orientation. Hence, this form of angular correlations constitutes a critical prediction of the statistical connectivity model, that, to our knowledge, has never been tested with experimental data.
%	
%
%
\section*{Results}
\subsection*{Dipole orientation correlation function in cat and primate RGC mosaics}
%
%
%%%%%%%%%%%%%%%%%%%%%%%%%%%%%%%%%%
%
\begin{figure}[!ht]
\begin{center}
\includegraphics[width=\linewidth]{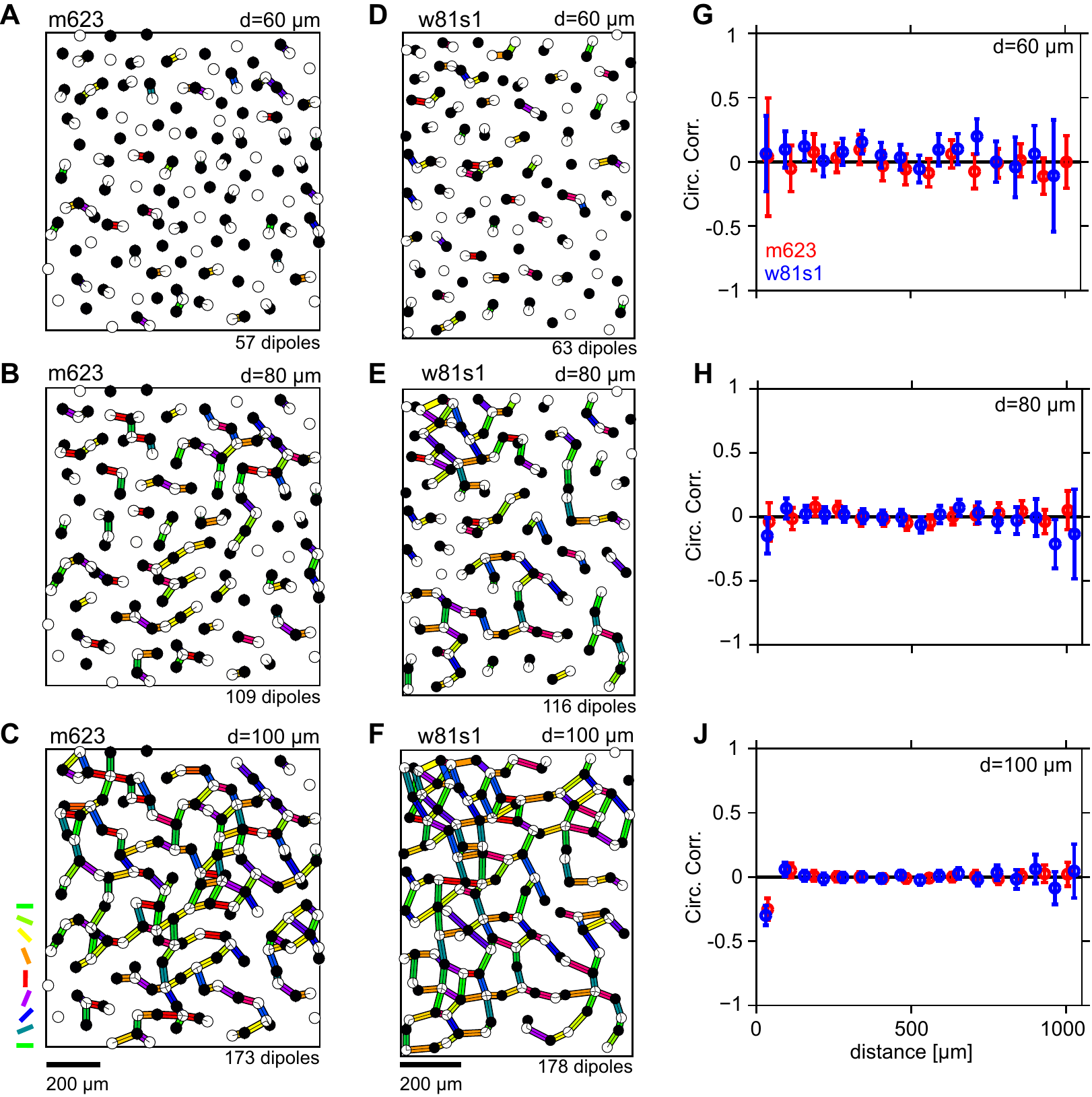}
\end{center}
\caption{\textbf{Spatial correlations of dipole orientations are absent in cat beta cell mosaics.}
\textbf{A} ON/OFF cells (empty/filled circles) for the cat beta cell mosaic m623 \cite{Zhan2000}. Preferred orientation of dipoles extracted for $d=60\mu m$ are shown as colored bars. Colorcode as in Fig. \ref{fig1}C.
\textbf{B} as A but for $d=80\mu m$. 
\textbf{C} as A but for $d=100\mu m$. 
\textbf{D}-\textbf{F} as A-C but for cat beta cell mosaic w81s1 \cite{Zhan2000}. 
\textbf{G} Correlation of dipole orientations for m623 (red) and  w81s1 (blue), calculated from dipoles extracted for $d=60\mu m$. Error bars indicate 95\% confidence intervals of bootstrap distributions.
\textbf{H} as G but for $d=80\mu m$. 
\textbf{J} as G but for $d=100\mu m$. 
\label{fig4}}
\end{figure}
We first determined the correlation function of dipole orientations in the two published cat beta cell mosaics. These mosaic fields will be referred to by their keys: w81s1 and m623. Field w81s1 was created by digitizing the map shown in Figure 6 of \cite{Wassle1981a}. Field m623 was taken from \cite{Zhan2000}. To limit the restriction of considering only nearest neighbor ON/OFF cells pairs, we followed the flexible dipole definition introduced in \cite{Paik2012b} (Fig. 1C). In this definition, a parameter $d$  is introduced, describing a distance below which neighboring pairs of ON/OFF cells are considered as dipoles (see Eq. (\ref{Eq:new_interaction_function})). The larger $d$, the more dipoles are formed by each RGC cell (cf. Fig. \ref{fig1}C). The nearest neighbor distance distribution of the mosaic defines a range of sensible $d$-values for each mosaic. For instance, the nearest neighbor distribution of cat beta cell mosaics peaks around $60\mu m$ \cite{Wassle1981a} and, therefore, $d$-values smaller than 60$\mu m$ lead to the extraction of only very few dipoles. On the other hand, values larger than 100$\mu m$ lead to many dipoles between non-nearest neighbors.\\ 
Figures \ref{fig4}A-C show the m623 mosaic along with all the dipole pairs, color-coded according to their orientation extracted for $d = 60\mu m$ (A), $d = 80\mu m$ (B),  and $d = 100\mu m$ (C) (see Materials and Methods). Figures \ref{fig4}D-F display the dipoles found for the mosaic w81s1. For $d = 80\mu m$ and $100\mu m$, dipoles in both mosaics appear as organized into linear chains which, at first sight, one might take as an indication of spatial correlation in the dipole orientations \cite{Paik2012b}. However, a quantitative analysis reveals that such correlations are absent for all d-values (Figs. \ref{fig4}G-J). The only statistically significant correlation present in both mosaics is a weak anti-correlation on very short scales ($<100\mu m$) for $d = 100\mu m$. This finding will be explained in more detail below.\\
%
%
%
%
%%%%%%%%%%%%%%%%%%%%%%%%%%%%%%%%%%
%
\begin{figure}[!ht]
\begin{center}
\includegraphics[width=10cm]{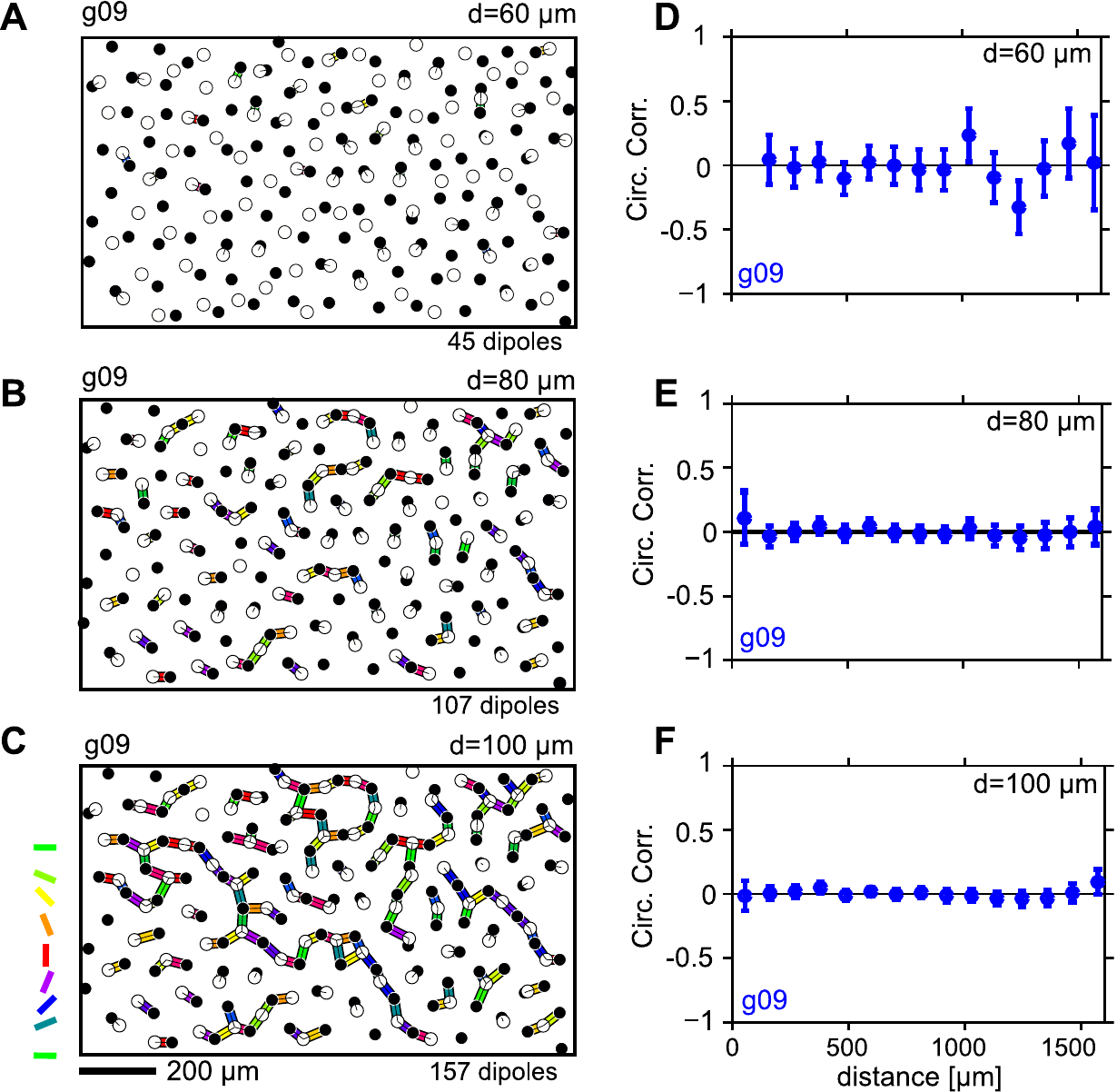}
\end{center}
\caption{\textbf{Spatial correlations of dipole orientations are absent in a primate parasol receptive field mosaic}
\textbf{A}  ON/OFF cells (empty/filled circles) for primate parasol cell receptive field mosaic G09 \cite{Gauthier2009}. Preferred orientation of dipoles extracted for $d=60\mu m$ are shown as colored bars. Colorcode as in Fig. \ref{fig1}C.
\textbf{B} as A but for $d=80\mu m$. 
\textbf{C} as A but for $d=100\mu m$. 
\textbf{D} Correlation of dipole orientations for mosaic G09, calculated from dipoles extracted for $d=60\mu m$. Error bars indicate 95\% confidence intervals of bootstrap distributions.
\textbf{E} as D but for $d=80\mu m$. 
\textbf{F} as D but for $d=100\mu m$. 
\label{fig_primates}}
 \end{figure}
%%%%%%%%%%%%%%%%%%%%%%%%%%%%%%%%%%%%%%
The two mosaics analyzed so far are based on the position of cell bodies and not those of the actual RGC receptive fields. Since these two do not match necessarily, it is important to test whether RGC receptive field mosaics substantially differ from cell position mosaics with respect to their dipole correlation structure. Therefore, we next repeated the above analysis for a previously published primate parasol cell receptive field mosaic (referred to by its key G09 \cite{Gauthier2009} in the following). Figures \ref{fig_primates}A-C show the G09 mosaic along with all the dipole pairs, color-coded according to their orientation extracted for $d = 60\mu m$ (A), $d = 80\mu m$ (B),  and $d = 100\mu m$ (C). Again, a quantitative analysis reveals that dipole angular correlations are absent for all d-values (Figs. \ref{fig_primates}D-F). \\
To analyze local correlations more systematically, we varied the parameter $d$ for all three mosaics in the data set over the entire range of sensible values and determined angular correlations in each of the mosaics in the first distance bin (see Materials \& Methods). Figures \ref{local_correlation_epiphenomenon}A and B depict the results of this analysis. Interestingly, for small $d$-values, all mosaics exhibit a weakly positive local correlation, whereas for larger d-values dipoles appear anti-correlated in all three mosaics (see Fig. \ref{fig4}J for an example). Intuitively, this dependency of correlation values can be understood is a consequence of our flexible dipole definition together with a typical distance between nearest neighbor RGC (Fig. \ref{local_correlation_epiphenomenon}C).  For small and intermediate $d$-values, an ON-cell positioned between several OFF-cells forms dipoles with mostly one or two of them. If dipoles are formed with two OFF cells, due to the regular spacing of OFF cells, the angles formed by these two dipoles are likely positively correlated (Fig. \ref{local_correlation_epiphenomenon}A, left inset). For larger $d$-values, more than two dipoles are typically formed. In this case, the regular spacing leads to an effective anti-correlation between their angles (Fig. \ref{local_correlation_epiphenomenon}A, right inset). 
To investigate whether such correlations suffice to seed the development of spatially repetitive and smooth OPMs in V1, we compared the correlation traces found in the experimentally measured mosaics to model mosaics obtained by a pairwise interacting point process (PIPP) \cite{Eglen2005, Hore2012}. The PIPP is a method for the generation of a spatial distribution of points specifying only pairwise interaction between individual points. It has been previously shown to accurately reproduce the spatial statistics of experimentally measured RGC mosaics \cite{Eglen2005, Hore2012}. With parameters fitted to experimental data, the PIPP generates regularly spaced RGC mosaics with radially isotropic autocorrelograms and lack of long-range positional order \cite{Hore2012}. For such mosaics, the statistical connectivity framework predicts OPMs that lack a typical column spacing and are qualitatively different from experiment \cite{Ringach2004a, Hore2012}. Figures \ref{local_correlation_epiphenomenon}D,E show that the local correlation of dipole angles in the model mosaics exhibit the same d-dependence as the data. For small $d$-values mosaics exhibit a weakly positive local correlation, whereas for larger $d$-values dipoles are weakly anti-correlated. We conclude that the weak local correlations found in experimental mosaics are a consequence of our flexible definition of RGC dipoles together with a typical spacing between neighboring cells. The quantitative match between PIPP mosaics and experimental data together with the fact that PIPP mosaics seed unrealistic cortical OPMs in the statistical connectivity model indicate that these correlations are not sufficient to seed the development of realistic cortical OPMs.
\subsection*{An OPM-modulated pairwise interacting point process}
%
%%%%%%%%%%%%%%%%%%%%%%%%%%%%%%%%%%
%
\begin{figure}[!ht]
\begin{center}
\includegraphics[width=14.0cm]{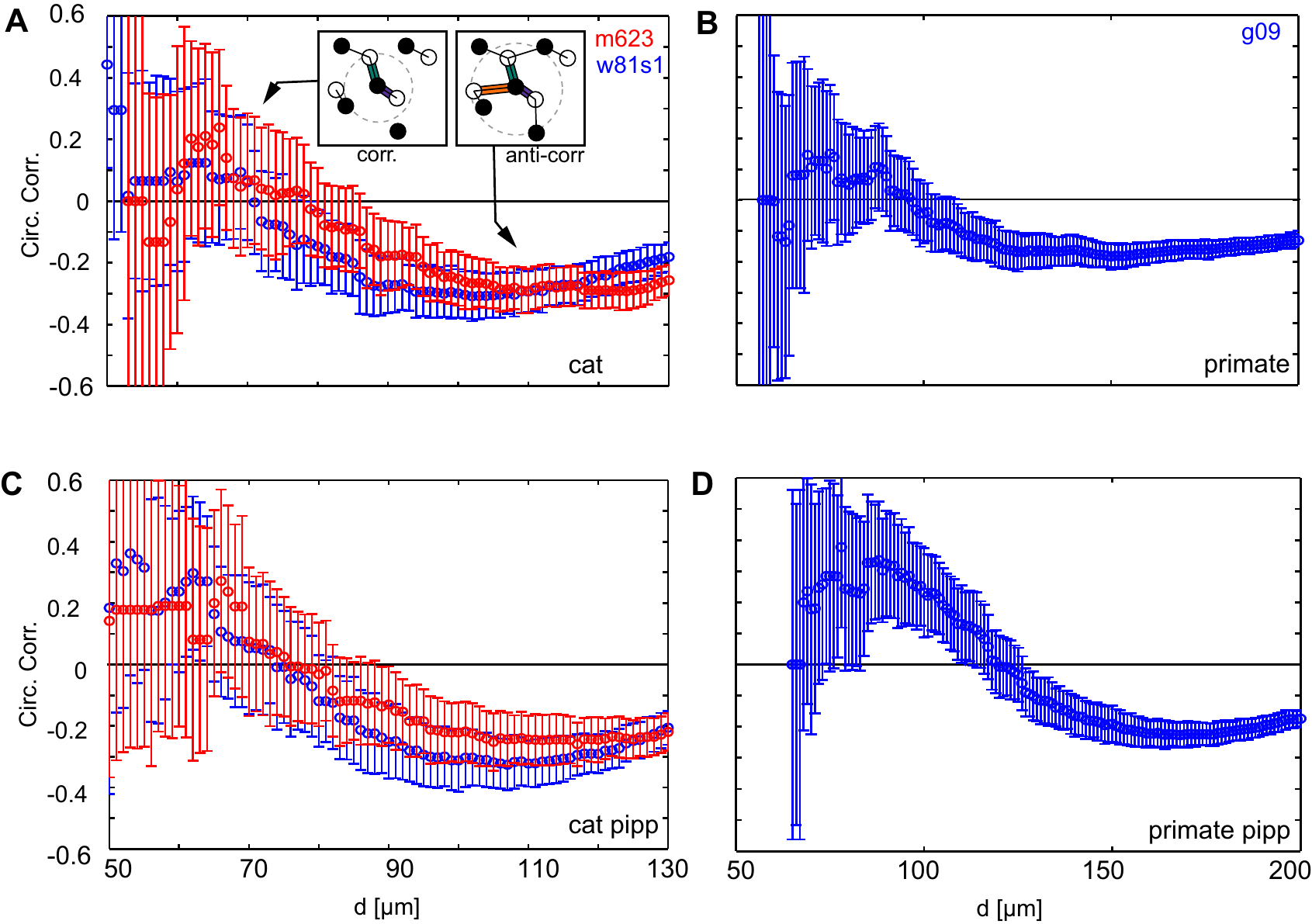}
\end{center}
\caption{\textbf{Measured local correlation values depend on choice of dipole distance parameter $d$.}
\textbf{A} Local correlations in cat beta cell mosaics (red: m623, blue: w81s1) as a function of the dipole extraction parameter $d$ (see Fig. \ref{fig1}C).
\textbf{B} As A but for primate parasol cell receptive field mosaic G09.
\textbf{C} Weak local positive or negative correlations emerge in the same RGC configuration, depending on dipole extraction parameter $d$.
\textbf{D} As A but for simulated PIPP mosaics with parameters fitted to cat beta cell mosaics m623 (red) and w81s1 (blue).
\textbf{E} As B but for simulated PIPP mosaics with parameters fitted to primate parasol receptive field mosaics G09. All error bars indicate 95\% confidence intervals of bootstrap distributions.
}
\label{local_correlation_epiphenomenon}
\end{figure}
One might wonder whether the absence of detectable positive or negative correlations in dipole angles is merely a consequence of the small sizes of each of these data sets.
In fact, all three mosaics analyzed each contain only about 100 cell positions (or RF center positions) and a similar number of dipoles. To clarify whether such small data sets are sufficient to detect both relevant dipole correlations, it is necessary to design model mosaics, e.g. defined by a some point process, with realistic spatial statistics and with a known degree of dipole angle correlations. Using such a point process, one can then ask whether the size of the available data sets in principle permits detection of such correlations. However, up to this point, a suitable point process with a known degree of dipole angle correlations has not been proposed. We now introduce a pairwise interacting point process (PIPP) with such characteristics, and start by briefly outlining the definition of the conventional PIPP for RGC mosaics as introduced in \cite{Eglen2005}.\\
The PIPP is a method for the generation of a spatial distribution of points specifying only pairwise interaction between individual points. Interactions between points are usually specified in pairwise interaction functions. The product of these pairwise interaction functions for a specific location for all possible pairs of points gives the probability of finding a point at a particular position. For bivariate data such as the positions $\mathbf{x}^i_\text{ON}$ and $\mathbf{x}^i_\text{OFF}$ of a mosaic of ON-cells and a mosaic of OFF-cells, the PIPP is characterized by two intra-mosaic interaction functions $h_{\text{ON},\text{ON}}(|\mathbf{x}^i_\text{ON}-\mathbf{x}^j_\text{ON}|)$, $h_{\text{OFF}, \text{OFF}} (|\mathbf{x}^i_\text{OFF}-\mathbf{x}^j_\text{OFF}|)$ and one inter-mosaic interaction $h_{\text{ON}, \text{OFF}}(|\mathbf{x}^i_\text{ON}-\mathbf{x}^j_\text{OFF}|)$. To describe the positioning of beta cells in the cat retina, Eglen et al. \cite{Eglen2005} used a parametric form of repulsion
\begin{eqnarray*}
  h(r) &=& \left\{
  \begin{array}{l l}
    0 & \quad \text{if $r\le\delta$}\\
    1-\exp\left(-\left|\frac{r-\delta}{\varphi}\right|^\alpha\right) & \quad \text{if $r>\delta$}\\
  \end{array} \right.
\end{eqnarray*}
for all three interaction functions, with $r = |\mathbf{x}^i_\text{ON}-\mathbf{x}^j_\text{ON}|$, $r = |\mathbf{x}^i_\text{OFF}-\mathbf{x}^j_\text{OFF}|$ or $r = |\mathbf{x}^i_\text{ON}-\mathbf{x}^j_\text{OFF}|$ for $h_{\text{ON}, \text{ON}}$,  $h_{\text{OFF}, \text{OFF}}$ or  $h_{\text{ON}, \text{OFF}}$, respectively. By fitting the parameters $\alpha$ and $\varphi$ to experimental data, they showed that inter-mosaic interactions are sufficiently described by solely ensuring that two cells are not less than the soma distance apart, i.e. 
\begin{eqnarray}
  h_{\text{ON}, \text{OFF}}(r) &=& \left\{
  \begin{array}{l l}
    0 & \,\text{if $r\leq\delta$}\\
    1 & \, \text{if $r>\delta$}\,,
  \end{array} \right.
   \label{Eq:old_interaction_function}
\
  \end{eqnarray}
with $\delta$ being the soma diameter. Intra-mosaic interactions were best fit by values for $\alpha$ and $\varphi$ which ensure a semiregular placement of RGC cells without long-range positional order \cite{Eglen2005}. Such mosaics lack non-local spatial order in their dipole angles and, hence, the statistical connectivity framework predicts orientation maps that lack a typical column spacing \cite{Ringach2004a,Hore2012}. \\
To introduce correlations of dipole angles into the PIPP, we start by formalizing the illustration in Fig. \ref{fig1}A. The dipole vector $\mathbf{x}^i_\text{ON}-\mathbf{x}^j_\text{OFF}$ between an ON/OFF pair of RGCs points in the direction $\arg(\mathbf{x}^i_\text{ON}-\mathbf{x}^j_\text{OFF})$ in the interval $[-\pi ,\pi)$.  The preferred orientation of the dipole (Fig. \ref{fig1}A, most right) can then be mathematically expressed as
\begin{equation}
\phi(\mathbf{x}^i_\text{ON},\mathbf{x}^j_\text{OFF}) = \text{mod} \left(\text{arg}(\mathbf{x}^i_\text{ON}-\mathbf{x}^j_\text{OFF}) + \pi/2 ,\pi \right) \,,
\label{Eq:dipole_def}
\end{equation}
varying in the interval $[0,\pi)$. The main idea is now to modify the inter-mosaic interaction function (Eq. (\ref{Eq:old_interaction_function})) to add a dipole correlational structure that matches the spatial correlations of orientation preferences in visual cortical OPMs. A visual cortical OPM can be represented as $\theta(\mathbf{x})$ with $\theta \in [0,\pi)$. Note that $\mathbf{x}$ here describes positions on the retina. An OPM measured with optical imaging of intrinsic signals \cite{Bonhoeffer1991,Blasdel1992a} is naturally given as $\theta'(\mathbf{X})$, where $\mathbf{X}$ specifies cortical position. The retinotopic map $\mathbf{x} = \mathbf{R}(\mathbf{X})$ associates a cortical position $\mathbf{X}$ with a position $\mathbf{x}$ on the retina. Via the inverse transformation, the representation of the OPM in retinal coordinates is obtained, i.e. $\theta(\mathbf{x}) \equiv \theta'(\mathbf{R}^{-1}(\mathbf{x}))$. In the following, only small subregions of OPMs were considered and the retinotopic map was assumed to be linear.\\
Using the OPM representation $\theta(\mathbf{x})$, we modify the inter-mosaic interaction function Eq. (\ref{Eq:old_interaction_function}) by multiplying it with a function $h_\gamma(\mathbf{x}^i_\text{ON},\mathbf{x}^j_\text{OFF})$ that depends on the difference between the preferred dipole angle $\phi(\mathbf{x}^i_\text{ON},\mathbf{x}^j_\text{OFF})$ and the preferred orientation specified in the OPM at position $(\mathbf{x}^i_\text{ON}+\mathbf{x}^j_\text{OFF})/2$ (half way between the two RGCs):
\begin{eqnarray}
H_{\text{ON},\text{OFF}}(\mathbf{x}^i_\text{ON},\mathbf{x}^j_\text{OFF})=  h_{\text{ON},\text{OFF}}(r)\cdot h_\gamma(\mathbf{x}^i_\text{ON},\mathbf{x}^j_\text{OFF})\,.
\label{eq:intra-interaction_function_definition}
\end{eqnarray}
We choose
\begin{eqnarray}
h_{\gamma}(\mathbf{x}^i_\text{ON},\mathbf{x}^j_\text{OFF}) &=& \left\{
  \begin{array}{l l}
    \exp\left\{\gamma \left[ \cos\left(\phi(\mathbf{x}^i_\text{ON},\mathbf{x}^j_\text{OFF}) - \theta\left((\mathbf{x}^i_\text{ON} + \mathbf{x}^j_\text{OFF})/2\right) \right) - 1\right] \right\} & \, \text{if $r\leq d$}\\
    1 & \, \text{if $r>d$}\,.\\
  \end{array} \right.
  \label{Eq:new_interaction_function}
\end{eqnarray}
In addition to ensuring that two cells are not less than the soma distance apart, this new inter-mosaic interaction function enforces that the positioning of an ON cell at position $\mathbf{x}^i_\text{ON}$ and an OFF cell at position $\mathbf{x}^j_\text{OFF}$ is such that the preferred orientation $\phi(\mathbf{x}^i_\text{ON},\mathbf{x}^j_\text{OFF})$  of the ON/OFF pair is similar to preferred orientation specified in the OPM at the corresponding point. In this way, the OPM $\theta(\mathbf{x})$ is expected to modulate the positioning of ON/OFF cells when numerically simulating the positioning of the RGC mosaics with a Monte-Carlo procedure \cite{Eglen2005}  (Fig. \ref{fig1}D) such that preferred orientations of dipoles align with preferred orientations given in the OPM. Note that the preferred orientation of a dipole is orthogonal to the dipole vector $\mathbf{x}^i_\text{ON}-\mathbf{x}^j_\text{OFF}$ (see Fig. \ref{fig1}A). With respect to the PIPP, this model has two additional parameters. The strength of modulation through the OPM is specified by a modulation parameter $\gamma$. If $\gamma$ is zero, the positioning of ON/OFF pairs is not influenced by the OPM region and the model is equivalent to the PIPP model. The larger the value of $\gamma$, the stronger the penalty for OPM and dipole angle differing. We refer to the process specified by this inter-mosaic interaction function as \textit{modulated PIPP} (mPIPP). The second parameter is the distance $d$ below which neighboring pairs of ON/OFF cells are considered as dipoles (see Eq. (\ref{Eq:new_interaction_function})). This parameter is taken to be the same as the parameter for defining dipoles (cf. Fig. \ref{fig1}C). Again, the larger $d$, the more dipoles each RGC cell is assumed to form with surrounding cells (cf. Fig. \ref{fig1}C). The nearest neighbor distance distributions of the different mosaics imply sensible values for the choice of $d$.\\
We would like to emphasize that by defining the mPIPP as above, we by no means imply any influence of cortical orientation preference upon the positions of ON/OFF RGCs during postnatal development. The mPIPP merely is a phenomenological algorithm to attempt to ``reverse engineer" one plausible realization of an RGC mosaic with the necessary spatial structure to yield an OPM with realistic spatial properties within the statistical connectivity model as considered in \cite{Paik2011}.
\subsection*{Statistical characterization of mPIPP mosaics}
%%%%%%%%%%%%%%%%%%% FIGURE %%%%%%%%%%%%%%%%%%
\begin{figure}[!ht]
\begin{center}
\includegraphics[width=14cm]{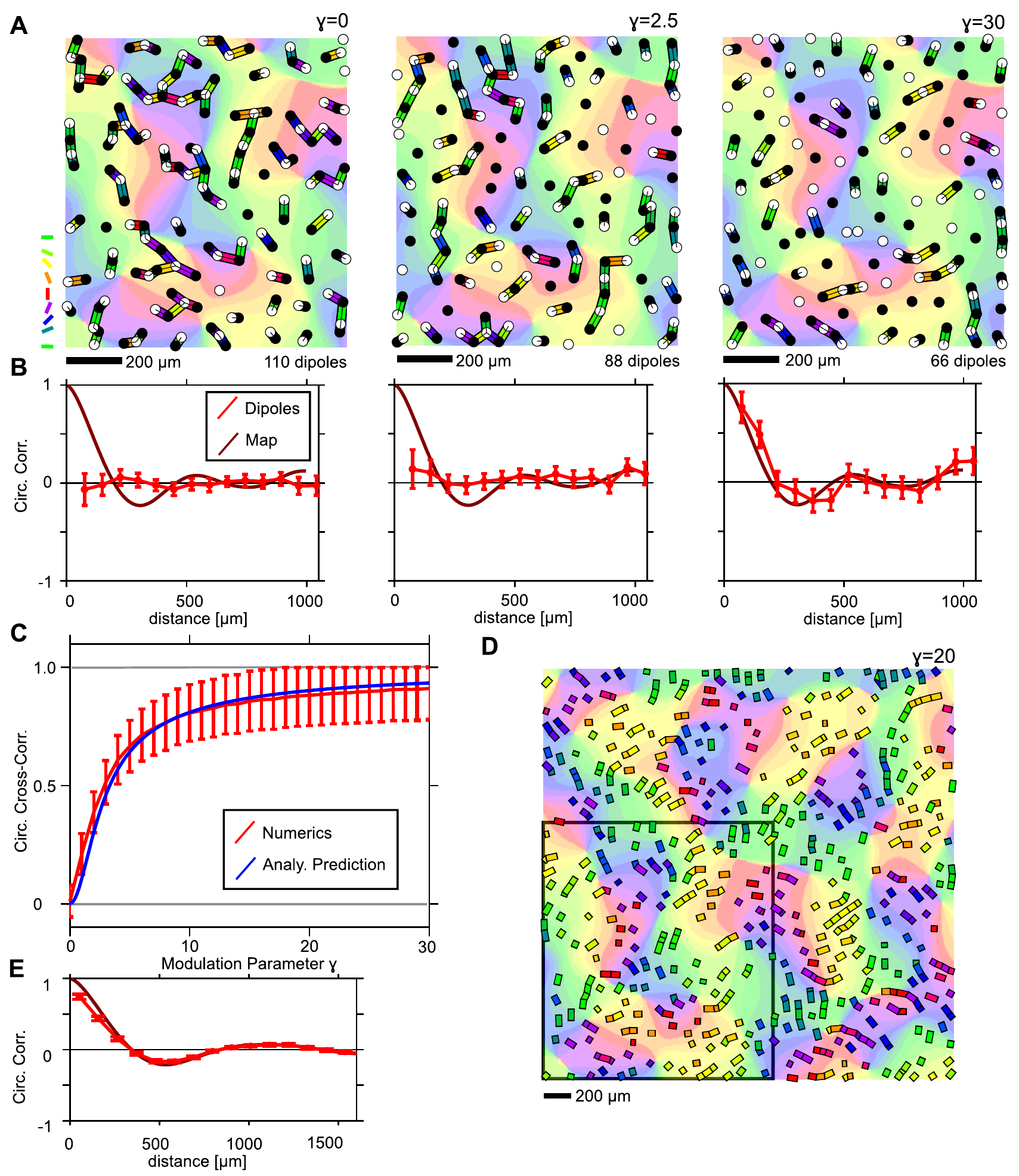}
\end{center}
\caption{\textbf{Statistical properties of mPIPP mosaics.}
\textbf{A} mPIPP RGC mosaics obtained with modulation parameter $\gamma = 0$ (left), $\gamma = 2.5$ (middle) and $\gamma = 30$ (right). ON (OFF) cells are displayed as empty (filled) circles. Dipoles are indicated as colored bars with colors indicating their preferred orientation.  OPM in the background is the region used to modulate the PIPP. Colorcodes as in Fig. \ref{fig1}C. Parameters for the PIPP were chosen as in \cite{Eglen2005} for cat beta cell mosaic m623 \cite{Zhan2000}. The cortical magnification is $\xi=1.7\text{mm}_c/\text{mm}_r$. Dipoles were extracted with $d=80\mu m$ (see Materials and Methods).
\textbf{B}  Correlation of dipole angles for the mosaics shown in A (red circles: $\gamma=0$ (left), 2.5 (middle), 30 (right)). Dark red line indicates correlation function of the modulating OPM region in A.
\textbf{C} Cross-correlation between dipole orientations and modulating OPM region for different values of the modulation parameter $\gamma$. Red line: numerical simulations. Blue line: analytical prediction (see Materials and Methods).
\textbf{D} RGC dipole pattern obtained for an mPIPP with $\gamma=20$ from a $3\Lambda\times3\Lambda$-OPM. All other parameters as in A. For clarity, only the dipoles formed by the mosaics are shown as colored rectangles. The modulating OPM region is shown in the background. Black square indicates the OPM region used to simulated the mPIPP in A.
\textbf{E} As B for the mosaic in D ($\gamma = 20$). Note, that the dipole correlation function closely follows the correlation function of the modulating OPM. All error bars indicate 95\% confidence intervals of bootstrap distributions.
\label{fig2} }
\end{figure}
To characterize the mPIPP defined above, we first investigated how the positioning of ON/OFF RGC dipoles is altered by the modulation through realistic OPMs. To this end, we extracted small regions of previously published OPMs from cat area 17 (data courtesy of Z. Kisvarday, see Materials and Methods) and used them to modulate the PIPP according to the above description. Rectangular OPM regions must be chosen such that the size of the retinal region corresponding to the OPM fits the two cat beta cell somata position mosaics w81s1 and m623 \cite{Wassle1981a,Zhan2000}. Thus, the size of the extracted cortical region depends on the cortical magnification factor $\text{mm}_c/\text{mm}_r$ at the position the mosaic has been recorded from. The larger the magnification factor, the larger the extracted cortical region has to be.\\
The center of mosaic w81s1 was located $19^\circ$ below the mid line of the visual streak, $4 mm$ from the area centralis \cite{Zhan2000}. The mosaic m623 has been obtained from 5mm eccentricity and $5.5^\circ$ below the mid line of the visual streak. Both fields are situated at positions in the visual field with similar cortical magnification factor (see Materials and Methods), estimated to be $\xi=1.7\text{mm}_c/\text{mm}_r$, where $\text{mm}_c$ is $\text{mm}$ on the cortex and $\text{mm}_r$ is millimeter on the retina. Both RGC fields are of similar linear extent ($\approx$1mm). Hence, extracted visual cortical regions were of  $\approx$1.7mm linear extent corresponding to an area of approximately 3 hypercolumns \cite{Kaschube2002,Kaschube2009}. \newline
Figure \ref{fig2}A show mosaics obtained by simulating an mPIPP with a Monte-Carlo procedure for a cortical magnification of $\xi=1.7\text{mm}_c/\text{mm}_r$, modulated by a cat OPM region for $\gamma=0$ (left), $\gamma = 2.5$ (middle) and $\gamma=30$ (right). ON/OFF pairs less than $d = 80\mu m$ apart were considered to form dipoles. All other parameters were chosen as in \cite{Eglen2005} for m623 (see also Suppl. Table 1). In the case $\gamma=0$ (original PIPP), the dipole orientations are not correlated with the preferred orientations specified in the OPM. For $\gamma=2.5$, some dipoles tend to locally align their orientation to match the orientation preference given in the OPM but most dipoles have random orientations. For $\gamma=30$, the dipole orientations are strongly correlated with preferred angles specified in the OPM. mPIPP realizations generally display fewer dipoles than the experimental mosaic m623 for the same $d$-values (cf. Fig. \ref{fig4}B). This is true even for the conventional PIPP process (m623: 109 dipoles; mPIPP $\gamma=0$: 98.6	 dipoles on average; $\gamma = 2.5$: 77.8 dipoles on average; $\gamma = 30$: 55.4 dipoles on average). This can be attributed to a slightly increased inhomogeneity of both the PIPP and mPIPP compared to the experimental data (see also Fig. 2 in \cite{Eglen2005}). Figure \ref{fig2}B shows the spatial correlation of dipole angles for all three $\gamma$-values (see Materials and Methods). With increasing $\gamma$, dipole orientations become locally correlated for distance smaller than $200\mu m$. Correlation drops to zero around $200\mu m$ distance and negative values are obtained between $200$ and $500\mu m$. Moreover, the correlation function of dipole orientations approaches the correlation function of the OPM for increasing $\gamma$ (solid dark red lines in Fig. \ref{fig2}B). Similarly, the cross-correlation between the preferred orientation given by the OPM and the orientation of the dipole rapidly increases with $\gamma$ (Fig. \ref{fig2}C). Finally, we considered mPIPP mosaics modulated by larger OPM regions (size 3x3 column spacings $\Lambda$,  Fig. \ref{fig2}D). Again, spatial correlations of preferred dipole orientations approach the correlation of preferred orientations in the OPM for increasing $\gamma$ (Fig. \ref{fig2}E). \\
The dipole angle correlation function periodically modulates around zero because in mPIPP mosaics dipole angles roughly repeat within a typical distance. This shows that the mPIPP can generate RGC mosaics that within the statistical connectivity framework predict orientation maps with a typical distance between adjacent column preferring the same orientation. The typical distance upon which dipole angles roughly repeat depends on the column spacing of the OPM used to modulate the PIPP as well as the assumed cortical magnification factor. Smaller column spacing or larger magnification factors will lead to a smaller scale of periodicity in the dipoles and vice versa. Importantly, however, dipole angle correlations decay to zero if OPM regions chosen contain more than 4$\Lambda^2$ even for large $\gamma$ (Fig. \ref{fig2}E). Hence, long-range spatial order in the dipole angles are absent in the mPIPP mosaics. This clearly distinguishes the mPIPP dipoles from a pattern of dipoles obtained by Moir\'{e} interference of two noisy hexagonal lattices where long-range order is expected \cite{Hore2012}. 
%
%
%
%
%%%%%%%%%%%%%%%%%%% FIGURE %%%%%%%%%%%%%%%%%%%
\begin{figure}
\begin{center}
\includegraphics[width=14cm]{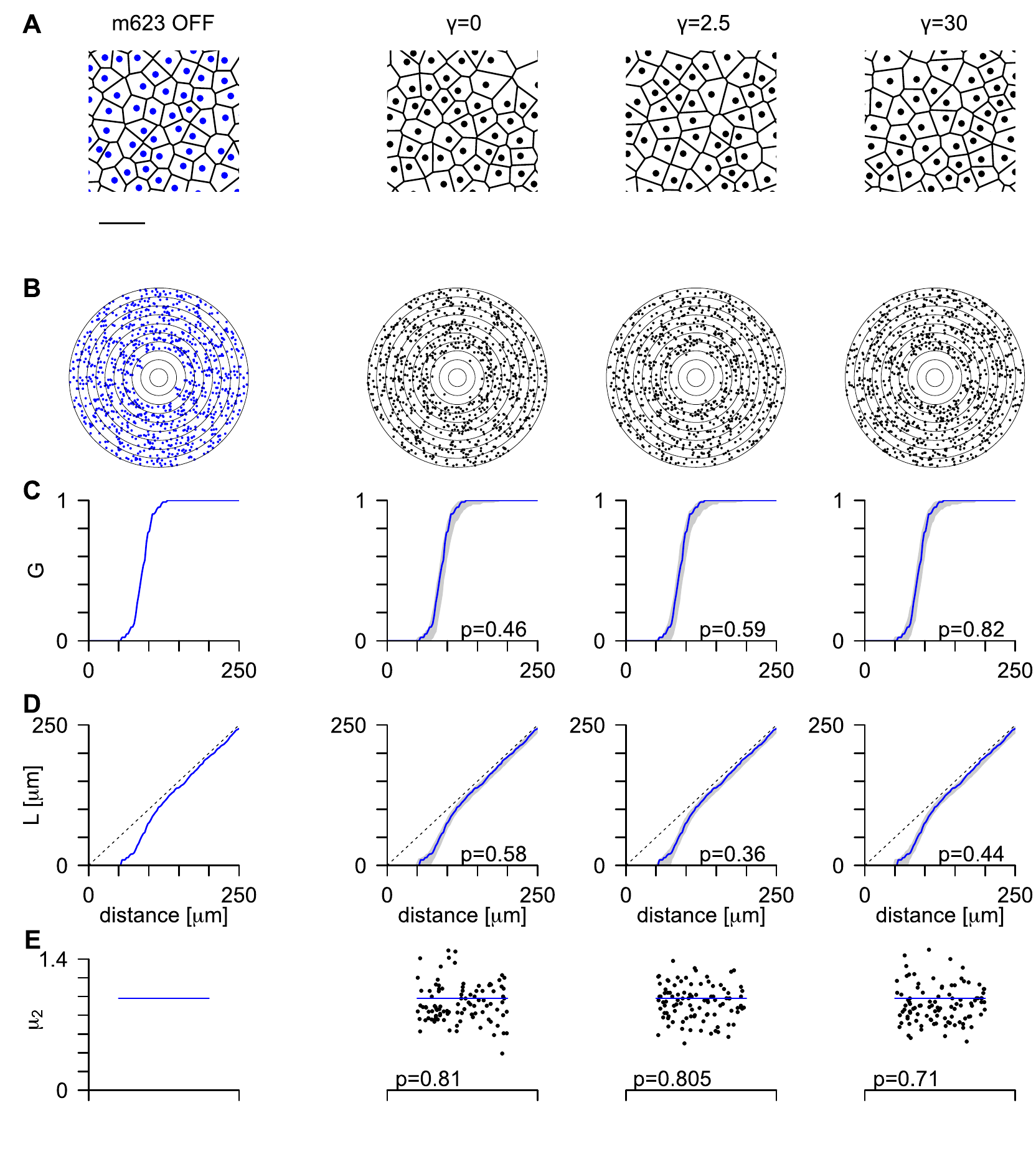}
\end{center}
\caption{\textbf{Spatial properties of cell positions in mPIPP mosaics are independent of $\gamma$ and in agreement with experimental data.}
Spatial statistics of the experimental OFF cells in mosaic m623 (column 1) and mosaics obtained by simulating an mPIPP model for three different $\gamma$ values (columns 2-4) and fixed $\xi=1.7\text{mm}_c/\text{mm}_r$ and $d=80\mu m$.
\textbf{A} Central region of the mosaics. Each point denotes one receptive field midpoint and is surrounded by its Voronoi polygon. Scale bar: 250 $\mu m$.
\textbf{B} Autocorrelogram of the points in A, with annuli drawn 25 $\mu m$ apart.
\textbf{C} Cumulative distribution of nearest neighbor distances ($G$-function, see Materials and Methods). The gray region shows the 95\% confidence interval of distributions from mPIPP simulations, and the solid line reflects the data (reprinted from column 1).
\textbf{D} $L$-functions for data and models, drawn in the same format as for panel C. The dashed line indicates the expectation for Poisson point process (complete spatial randomness), $L(t) \sim t$. \textbf{E} Topological disorder parameter $\mu_2$ for 99 realizations of 
the mPIPP (black dots) and for the data (horizontal blue 
line).
\label{eglen_stat_analysis_off}}
\end{figure}
%
%
%
%%%%%%%%%%%%%%%%%%%% FIGURE %%%%%%%%%%%%%%%%%%
%
%
\newline
We next wondered whether the short-range spatial statistics of ON/OFF cell positions was affected by placing dipoles such that their preferred orientation tends to align with realistic OPMs. To answer this question, we compared the nearest neighbor distance distributions, autocorrelograms of cell positions as well as spatial regularity of mPIPP mosaics with the data using statistical measures defined in \cite{Eglen2005, Hore2012} (see Materials and Methods). Figure \ref{eglen_stat_analysis_off} depicts the results of this analysis for $\gamma=0;2.5;30$ for the m623 OFF cell mosaic. Voronoi polygons and autocorrelograms of all simulated mPIPP mosaics bear close resemblance to the data (Figs. \ref{eglen_stat_analysis_off}A and B). The nearest neighbor statistics of all three mPIPP mosaics are statistically indistinguishable from each other and all three are statistically indistinguishable from the m623 field (Figs. \ref{eglen_stat_analysis_off}C and D). Figure \ref{eglen_stat_analysis_off}E plots a topological disorder parameter \cite{Kram2010,Hore2012}, $\mu_2$, measuring the spread of the distribution of the number of sides in each Voronoi polygon (Fig. \ref{eglen_stat_analysis_off}A, see Materials and Methods). $\mu_2$ is plotted for 99 simulated mosaics as dots along with the value obtained from the data as a blue horizontal line. For each value of $\gamma$ this line falls within the distribution of simulated values, indicating a good fit between mPIPP model mosaics and the data. Similar results were obtained when considering the w81s1 mosaic (data not shown). We conclude that the local statistics of cell positions is independent of the modulation parameter $\gamma$ and matches the statistics of experimentally observed mosaics.\newline
In summary, the proposed modulated PIPP generates semi-regular mosaics of ON and OFF center cells with realistic spatial statistics that within the statistical connectivity model predict experimentally observed OPMs. For increasing $\gamma$ the local spatial correlation between dipole orientations increases yet long-range positional order remains absent in the generated mosaics (Fig. \ref{fig2}). By varying a modulation parameter $\gamma$, we are able to tune cross-correlations between dipole orientations in the RGC mosaics and the preferred orientations given in the modulating regions of OPMs. The spatial statistics of ON/OFF cells positions, however, is unaffected by this modulation and all $\gamma$-values yield mosaics with spatial statistics consistent with the data.
\subsection*{Using mPIPP mosaics to assess the statistical power of correlation analysis}
%%%%%%%%%%%%%%%%%%%%%%%%%%%%%%%%%%%%%%%%%%%%%%%%
%
%
\begin{figure*}
\begin{center}
\includegraphics[width=14cm]{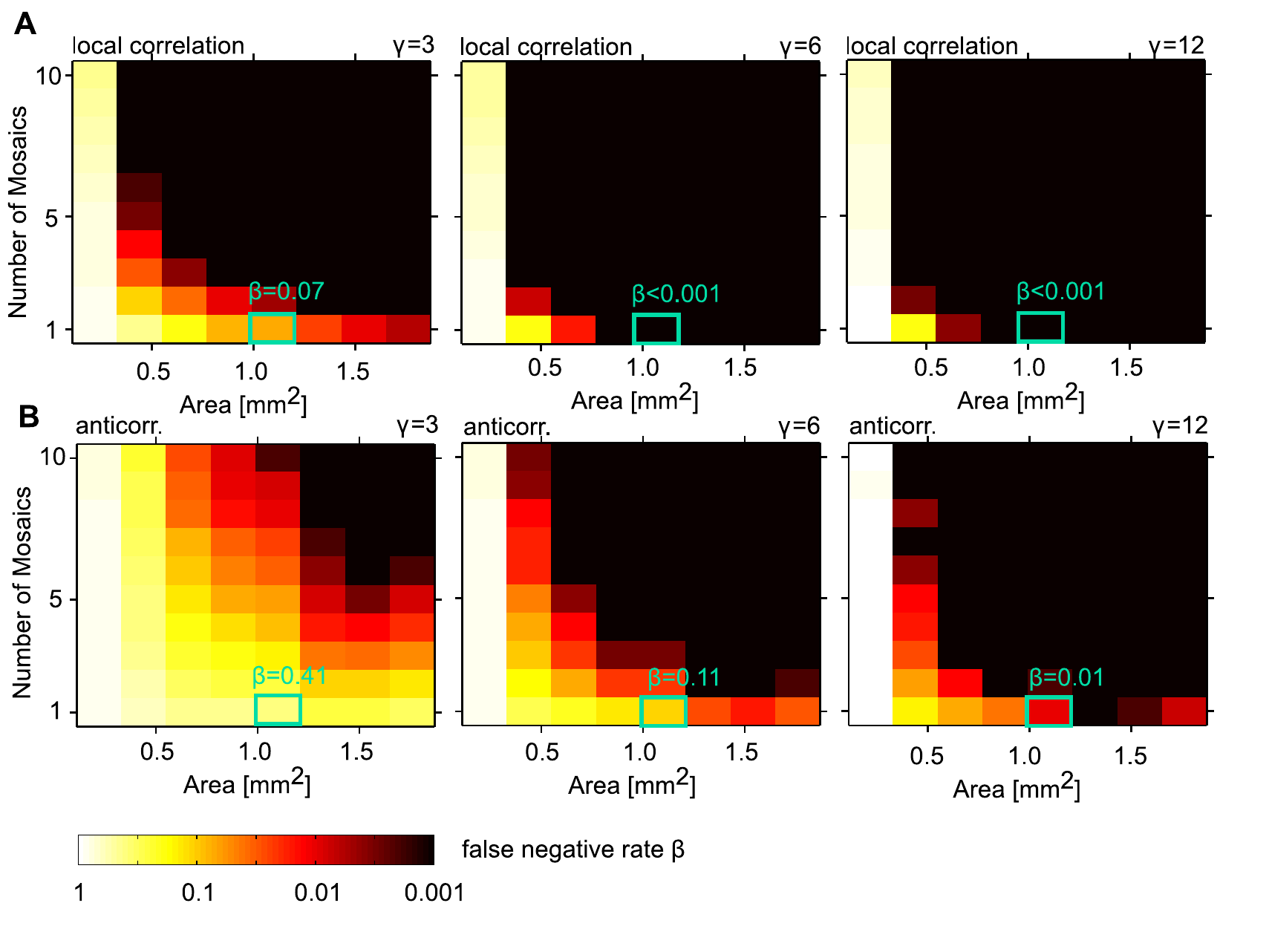}
\end{center}
\caption{\textbf{Estimating the statistical power of the test for the presence of spatial correlations in RGC dipole angles}
\textbf{A}  False negative rate (probability of failing to detect the positive local correlation) for mPIPP mosaics simulated with different modulation parameters $\gamma$ as a function of the number of mosaics N and their area size (see Materials \& Methods for details). Green box indicates size of cat RGC mosaic data sets analyzed in the present study (N = 1, area size $\approx$ 1mm$^2$). Note that even for $\gamma = 3$, the false negative rate with this data set size is very small.
\textbf{B} False negative rate for detecting negative correlation of dipole angles mPIPP mosaics around a distance of $300\mu m$ (see Fig. \ref{fig2}, see Materials \& Methods for details). In all panels, a cortical magnification of $\xi=1.7\text{mm}_c/\text{mm}_r$ was assumed and PIPP parameters were taken from the fit to the m623 mosaic \cite{Eglen2005}. 
\label{fig:error_II_test}}
\end{figure*}
Having characterized its statistical properties, we used the mPIPP to estimate the size of a data set necessary to detect correlations between dipole angles that could seed the emergence of locally smooth and repetitive OPMs. More specifically, we wanted to estimate the so-called false negative rate of a test for dipole angle correlations, i.e. the rate of rejecting the presence of correlations when correlations of a certain degree are present in the ensemble the data is drawn from. In principle, testing the presence of positive local dipole angle correlations will require fewer and smaller mosaics than the detection of presumably smaller negative correlations on larger spatial scales. To estimate the necessary size of a data set, we generated a number of N mPIPP mosaics with known degree of spatial correlation (cf. above) and varied their spatial extent. We then statistically compared these ensembles to N conventional PIPP mosaics of the same spatial extents \cite{Eglen2005} (see Materials \& Methods for details). Since PIPP mosaics lack the spatial dipole structure necessary for seeding cortical OPMs, this comparison can be employed to estimate a lower bound for detectable relevant dipole angle correlations, given a data set with N mosaics of fixed size, all measured at similar eccentricity. The latter implies that correlations functions of the mosaics can be averaged to improve the statistical power of the test. The two cat beta cell mosaics come from different eccentricities. Therefore, their correlation functions cannot be averaged.\\
Figure \ref{fig:error_II_test} depicts the estimates of the false negative error rate $\beta$ of the statistical test for the presence of dipole angle correlations. Figure \ref{fig:error_II_test}A shows the false negative rate for the detection of positive local correlation in mPIPP mosaics simulated with different modulation parameters $\gamma$ as a function of the number of mosaics and their area. Even for small $\gamma$, already one mPIPP mosaic of an area of 1mm$^2$ (the size of the cat beta cell data sets) yields $\beta=0.07$. This means that 93\% of the realizations of such mosaics are statistically distinguishable in terms of local dipole angle correlations from an ensemble of PIPPs. Similarly, 5 mosaics with an area of around 0.4mm$^2$ would be sufficient to reliably detect even weak local correlations  ($\gamma = 3$). Figure \ref{fig:error_II_test}B displays the probability of failing to detect negative correlation of dipole angles around a distance of $300\mu m$ in the mPIPP mosaics. As expected, averaging over more mosaics or, alternatively one larger mosaic is needed to detect such anti-correlations. From the above analysis, we conclude that both cat beta cell mosaics analyzed here are sufficiently large to detect even weak local positive correlations of dipole angles that could seed iso-orientation domains in V1, if present. However, a larger data set (e.g. $N>9$ mosaics of 1mm$^2$ measured at the same eccentricity) is needed to reliably test for the negative correlations that would be indicative of seeding spatially repetitive OPMs. Note that, since the mPIPP mosaics generally contain slightly fewer dipoles than experimental RGC mosaics (cf. Fig. \ref{fig2}A), all of the above estimates of statistical power are conservative and most likely even fewer or smaller samples of real mosaics would suffice to detect the presence of the respective correlations.
\subsection*{Comparison between mPIPP mosaics and cat beta-cell mosaics}
%%%%%%%%%%%%%%%%%%%%%%%%%%%%%%%%%%%%%%%%%%%%%%%%
\begin{figure}
\begin{center}
\includegraphics[width=\linewidth]{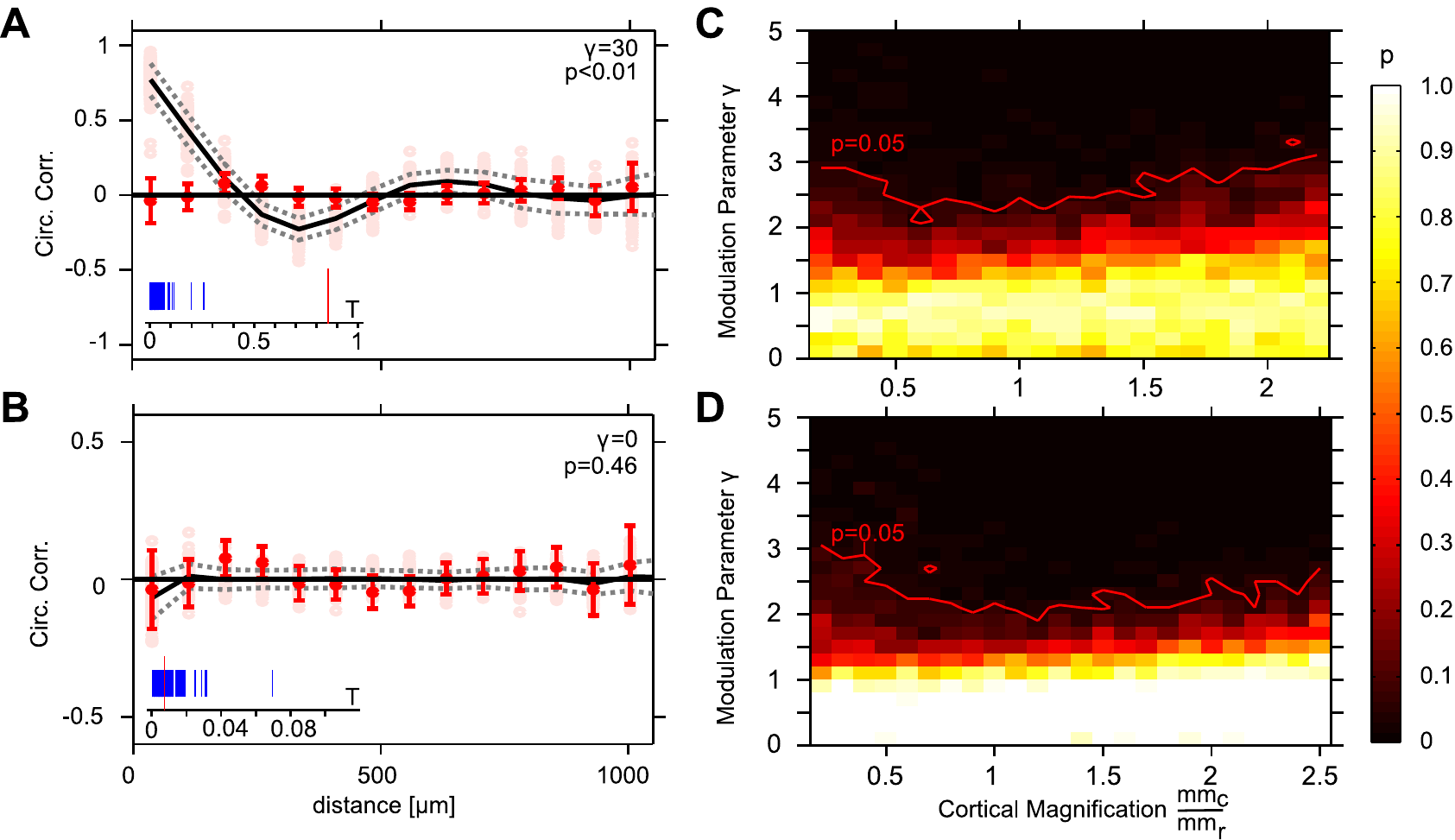}
\end{center}
\caption{\textbf{Constraining the modulation parameter $\gamma$ with experimental data.} \textbf{A} Correlation functions of dipole preferred orientations for 100 mPIPP realizations (pale pink dots, $\gamma = 30$, cortical magnification $\xi=1.7\text{mm}_c/\text{mm}_r$, $d=80\mu m$). Black drawn line indicates average correlation function, dashed lines show $\pm1\sigma$ deviation from the mean. Red dots indicate correlation function of dipole orientations for the mosaic m623 (redrawn from Fig. \ref{fig4}H). Insets show the T-distribution for Monte-Carlo data (blue) to estimate the p-value for the observed value (red) (see Materials and Methods). Note the correlation-anti-correlation-correlation structure of the correlation function. \textbf{B} As A but for $\gamma = 0$. \textbf{C} Monte-Carlo p-values (see Materials and Methods) for different values of $\gamma$ and the cortical magnification factor $\xi$ for m623. Red line indicates $p=0.05$ significance value.
\textbf{D} as C but for mosaic w81s1.
\label{fig5}}
\end{figure}
%
%
%
%%%%%%%%%%%%%%%%%%%%%%%%%%%%%%%
The absence of spatial correlations in dipole orientations in both published cat RGC mosaics can be employed to determine an upper bound for the modulation parameter $\gamma$ of the mPIPP. To do so, we synthesized mPIPP mosaics for a wide range of cortical magnification factors $\xi$ between $0.2\frac{ \text{mm}_c}{\text{mm}_r}$ and $2.5\frac{\text{mm}_c}{\text{mm}_r}$ (estimated value for both mosaics is $\xi=1.7$)  and for a wide range of modulation parameters $\gamma$ between 0 and 30. The maximum dipole distance was fixed at $d=80\mu m$. However, similar results were obtained with $d=60\mu m$ and $d=100\mu m$. Each mPIPP mosaic realization was modulated by a randomly chosen rectangular OPM region from cat area 17. The size of the OPM region was determined by the respective magnification factor $\xi$ (see Materials and Methods). For each pair of values ($\xi$, $\gamma$), we generated 100 mPIPP realizations and calculated their dipole orientation correlation functions. Figures \ref{fig5}A and B depict these correlation functions for $\xi=1.7\text{mm}_c / \text{mm}_r$ and $\gamma=30$ (A) and $\gamma = 0$ (B). For $\gamma = 0$, dipole orientation correlations of mPIPP mosaics and the experimentally measured mosaic perfectly overlap. For $\gamma = 30$, correlation values of experimental data lie far outside the min/max range of correlation values obtained with the mPIPP mosaics. To determine the range of $\gamma$ for which mPIPP mosaics are consistent with the data, we then asked for each pair of values ($\xi$, $\gamma$), how likely the correlation values in the data for distances smaller than $200\mu m$ are generated by the corresponding mPIPPs. For each pair ($\xi$, $\gamma$) this resulted in a Monte-Carlo p-value, indicating the likelihood of finding the correlation value of the experimental mosaic in the ensemble of mPIPPs. The p-values for different ($\xi$, $\gamma$) are depicted in Figures \ref{fig5}C (m623) and D (w81s1). For both mosaics, $p<0.05$ if $\gamma > 3$, indicating that the experimental data is only consistent with very small values of $\gamma$. For these values, positioning of RGC cells in the mPIPP is only very weakly modulated by the OPM region (see Figs. \ref{fig2}A and B). In particular, the data from w81s1 is most consistent with $\gamma=0$, i.e. the previously described PIPP \cite{Eglen2005} with complete absence of modulation from the OPM region. As emphasized before, in this regime the V1 OPM predicted by the statistical connectivity model does not exhibit a typical column spacing and therefore the model cannot account for the experimentally observed semiregular structure of visual cortical orientation maps.
\section*{Discussion}
In this paper we tested a fundamental prediction of the statistical connectivity model \cite{Soodak1986, Soodak1987, Ringach2004a, Ringach2007, Paik2011,Paik2012}. When considered in the so-called dipole approximation, i.e. when each V1 neurons receives input from only a few number of RGCs, the statistical connectivity model predicts that RGC dipole angles exhibit spatial correlations similar to the spatial correlations of orientation preferences in V1 OPMs. This means that dipole angles should be locally positively correlated and anti-correlated on larger scales. We analyzed two cat beta cell mosaics as well as one primate parasol receptive field mosaic searching for the presence of such correlations. All three mosaics lack spatial correlations on the relevant spatial scales (Figs. \ref{fig4}, \ref{fig_primates}). Weak local correlations can be attributed to receptive fields being semi regularly spaced and are not sufficient to seed realistic cortical OPMs (Fig. \ref{local_correlation_epiphenomenon}). \\
To investigate whether the absence of detectable correlations was merely a consequence of the small size of the available mosaic data sets, we used a novel point process (mPIPP) which generates realizations of an ON/OFF ganglion cell mosaic that could seed realistic OPMs \cite{Ringach2004a, Ringach2007, Paik2011}. By modulating the positioning of ON/OFF cells with experimentally obtained OPMs, the mPIPP generates semi-regular ganglion cell mosaics which within the statistical connectivity model will generate realistic spatially repetitive OPMs. Notably, no Moir\'{e}-Interference mechanism needs to be invoked to generate periodicity of the OPMs. The process extends previous work on PIPPs \cite{Eglen2005,Hore2012} to generate ON/OFF cell mosaics with realistic spatial properties. By varying a modulation parameter $\gamma$, the influence of the OPM region on the ON/OFF cells' position could be arbitrarily and predictably tuned (Fig. \ref{fig2}). The local spatial statistics of the model mosaics agree well with experiment (Fig. \ref{eglen_stat_analysis_off}). They are essentially insensitive to the modulation through the OPM. However, the inferred mosaics are characterized by a salient local correlation between neighboring dipole orientations, anti-correlation at intermediate distances and correlation at greater distances (Fig. \ref{fig2}), an effect of the typical distance between adjacent columns preferring the same orientation in the OPM.\\
The mPIPP mosaics were then used as reference mosaics with a predictable degree of dipole correlations to determine the statistical power of inferring the presence/absence of correlations from finite samples of RGC mosaics. We find that both cat beta cell mosaics are sufficiently large to reliably detect even weak local positive correlations of dipole angles that could seed iso-orientation domains of the size observed in experimental OPMs (Fig. \ref{fig:error_II_test}). However, a larger data set of mosaics measured at the same eccentricity would be needed to reliably detect the anti-correlations on larger spatial scales that would be indicative of seeding spatially repetitive OPMs. These findings then prompted us determine an upper bound for the local dipole angle correlations in the data (Fig. \ref{fig5}). Our results show that even weak correlations are ruled out by the data. Thus, experimentally measured RGC mosaics lack the local dipole structure to seed iso-orientation domains of the size observed in experimental OPMs.
\newline
As explained above the data sets analyzed in the present study are not sufficiently large to rule out a weak periodicity of dipole angles in experimentally measured RGC mosaics. However, we would like to emphasize that the presence of local correlations in ON/OFF dipole angles appears as a necessary prerequisite for the statistical connectivity model to yield spatially repetitive OPMs. When hexagonal RGC mosaics are considered, dipole angle correlations emerge via Moir\'{e} Interference of hexagonal mosaics, that are \textit{positionally} ordered over long distance ($> 1mm$) \cite{Paik2011, Paik2012}. Our algorithm to ``reverse engineer" plausible realizations of RGCs from measured OPMs shows that the statistical connectivity model does not necessarily rely on such long-range \textit{positional} order. In the PIPP model, correlations of neighboring ON/OFF dipole angles could instead be build into the mosaics by a simple modification of the intra-mosaic interaction function studied in \cite{Eglen2005}. The absence of angular correlations in the data not only constrains the modulation parameter $\gamma$ of the mPIPP to very small values. It also provides further evidence against the seeding of cortical OPMs via feed-forward projection of a Moir\'{e} interference pattern from the retina. \\
One main simplification of the statistical connectivity approach is that the transformation of visual inputs by the LGN is usually ignored. RGC afferent terminal axons diverge in the A-lamina, providing a one-to-many mapping between one X-RGC axon and several X-type geniculate relay cells in cat LGN. Moreover, geniculate neurons often receive input from several retinal afferents \cite{Usrey1999}. However, these multiple retinal inputs have mostly overlapping receptive-field centers \cite{Usrey1999}. This has important implications for how the pattern of retinal dipoles might be transformed into a pattern of LGN dipoles. The increased density of LGN dipoles might not interpolate the sparsely sampled dipole pattern of the retina. Therefore, the size of the iso-orientation domains set by the LGN dipoles does not increase with respect to the iso-orientation domains set by the RGCs.  Instead domains of dipoles of equal orientation sharply confined around RGC dipoles (similar to a Voronoi tessellation of the dipole center positions) might be the most likely outcome of the thalamic transformation. If so, the thalamic transformation would not be able to transform the uncorrelated dipole pattern in the retina into a pattern with angular correlations on the relevant scales.\\
The statistical connectivity model has been advanced as a theory explaining the establishment of a blueprint for the V1 orientation preference map during early visual development based on retinal inputs from the contralateral eye \cite{Paik2011}. In fact, the spatial layout of initially established contralateral eye dominated OPM is similar to the binocular OPM of the mature animal \cite{Crair1998}. Hence, retinal inputs from the contralateral eye alone should be sufficiently structured to generate a spatially repetitive OPM with typical column spacing in the millimeter range. These findings usually justify the analysis of properties of mosaics from one eye only while, of course, in the adult most visual cortical neurons are binocular \cite{Hubel1962}. If, as our analysis suggests, mosaics in the individual eyes lack the spatial structure necessary to seed realistic OPMs, it appears unlikely to us that by combining inputs from ipsilateral and contralateral eye, a correctly spatially structured seed could be established. To establish the necessary spatial structure from two spatially unstructured inputs, a considerable fraction of  V1 neurons would have to receive their OFF subfield input from one eye and their ON subfield input from the other. If this were the case, the OPMs measured by stimulating the ipsilateral or contralateral eye only in the adult animal should be considerably different. This is, however, not what Crair et al. have shown \cite{Crair1998}. We conclude that competing eye inputs are unlikely to offer an alternative non-cortical mechanism for generating the structure of experimentally measured OPMs.\\
The hypothesis (in this study called statistical connectivity model) that cortical OPMs could emerge from a spatially structured retinal organization has been considered with three different RGC mosaic classes and in two different regimes in terms of the number of feed-forward retinal inputs $N$ that a V1 neuron samples from. Both, Soodak and Ringach considered noisy hexagonal RGC mosaics and a large number of inputs to each cortical V1 cell ($N \approx$ 20) \cite{Soodak1987, Ringach2004a}. In this setup, the statistical wiring model results in OPMs without a typical column spacing and, hence, cannot account for the spatial structure observed in experimentally measured OPMs. Paik \& Ringach considered the model with noisy hexagonal RGC mosaics in the so-called dipole approximation, where input to a V1 cell is dominated by on average a close-by pair of one ON and one OFF cell \cite{Paik2011, Paik2012}. In this regime, OPMs predicted by the model exhibit a typical spacing between adjacent columns. However, as outlined in the introduction, the positional statistics of RGC mosaics is not well described by noisy hexagonal lattices \cite{Hore2012} and the PIPP mosaics provide a much better fit to the data. The statistical connectivity model considered with these more realistic mosaics fails to generate maps with a typical column spacing \cite{Ringach2007, Hore2012}. Our study introduces a third class of RGC mosaics (mPIPP mosaics) with realistic positional statistics \textit{and} a repetitive structure of cortical OPMs. The improved match between data and model, however, comes at the natural expense of spatial correlations in dipole angles in the mPIPP mosaics which are not found in experimentally measured mosaics.\\
The statistical connectivity model's mismatch with essential biological facts in each of these parameter regimes suggests that the spatial layout of OPMs is not determined by the structure of RGC mosaics and may instead result from intracortical mechanisms. One of the most striking demonstrations of how such intracortical mechanisms can shape visual cortex architecture comes from cat primary visual cortex. Using pharmacological treatments, Hensch and Stryker \cite{Hensch2004} locally altered the balance between intracortical inhibition and excitation during OPM formation. Enhancement of inhibitory circuits locally widened column spacing in V1 while local reduction of inhibition broadened the spacing of columns. While those findings are difficult to reconcile with the statistical connectivity hypothesis, theories in which cortical columns arise from an intracortical interplay between inhibition and excitation, e.g. \cite{vonderMalsburg1973,Swindale1982a, Wolf1998, Wolf2005}, could provide simple and plausible explanations for such an effect. Similarly, the progressive interareal and interhemispheric matching of features of columnar architecture such as the local column spacing in cat V1 suggests a strong influence of activity-dependent intracortical and even interareal interactions during postnatal column formation and critical period refinement and reorganization \cite{Kaschube2009} (see also \cite{Keil2010}). \\
The fact that RGC mosaics lack the spatial structure to seed realistic cortical OPMs does \text{a priori} not rule out the possibility that retinal/thalamic receptive field mosaics might exert an influence on the layout of OPMs during postnatal development or even provide a rough blueprint of visual cortical maps.  In fact, the representations of retinal blood vessel angioscotomas in the visual cortex in some squirrel monkeys are a striking demonstration of the influence of retinal features on cortical selectivity layouts \cite{Adams2002} (see \cite{Giacomantonio2007} for a theoretical treatment in terms of activity-dependent mechanisms). If this is the case, preferred orientations of RGC dipoles on the retina might still be a fairly good predictor of cortical orientation preference for a substantial fraction of neurons, and intracortical network interactions might serve to organize a retinal seed of orientation preference into a pattern with a typical spacing and a semi regular arrangement of pinwheel centers. Finally, it remains possible that a higher order statistical structure of RGC mosaics beyond simple dipoles (or their thalamic transformation) is capable of driving the formation of realistic OPMs. Including retinal biases or constraints into existing models for the activity-dependent self-organization of visual cortical orientation preference is needed to elucidate the interplay between subcortical feed-forward constraints and intracortical network self-organization. Along the same lines, experiments investigating the relationship between the structure of the RGC mosaics and the OPM in the same individual constitute a critical next step for future research.
\section*{Materials and Methods}
\subsection*{Optical imaging data \& preprocessing}
Optical imaging data from twelve adult cat area 17 hemispheres were used in this study. No experiments were carried out for the sole purpose of the present study. We reanalyzed partially published data, collected by the laboratory of Z. Kisv\'arday at Ruhr University Bochum.  The original animal license for these experiments was issued to Prof. Ulf Eysel (Dept. Neurophysiology)  and the research program was supported by SFB509. All experiments were conducted according to ethical regulations issued by the Ruhr University Bochum and conformed to the guidelines of the European Communities Council Directives, 1986 (86/609/EEC) as well as the German Animal Welfare Act. All animals derived from in house animal farm or from registered  breeders for experimental animals. \\
Surgery and preparation protocols have been described in detail elsewhere \cite{Buzas1998, Yousef1999}. Briefly, optical imaging of intrinsic signals were conducted on anesthetized (initial anesthesia: a mixture of ketamine, 7 mg/kg Ketanest (Parke-Davis, Berlin, Germany), and Xylazine, 1 mg/kg (Rompun, Bayer Belgium, Sint-Truiden, Belgium), i.m.; prolonged anesthesia: 0.4 -- 0.6\% halothane in a 1:2 mixture of O$_2$ and N$_2$O using artificial ventilation) and paralyzed (alcuronium chloride (0.15 mg/kg/h, Alloferin, Hoffman-La Roche, Grenzach-Whylen, Germany, i.a.) animals using the imaging system Imager 2001 (Optical Imaging, Germantown, NY) and the data acquisition software VDAQ (Version No. VDAQ218k, Optical Imaging). A craniotomy was performed on one hemisphere between stereotaxic coordinates (Horsley--Clarke) P7--A12 and L0.5--L6.5 to expose the cortical region corresponding to the representation of the central and lower parts of the visual field in both area 17 and area 18 \cite{Tusa1978}. 
Animals were monitored continuously throughout all procedures to ensure that adequate anesthesia was maintained. Area 17 identification was made on the basis of stereotaxic coordinates. Before acquiring data, the camera was focused at 650 -- 750$\mu$m below the cortical surface that was illuminated with 609$\pm$5 nm light. Visual stimuli were presented to one eye. Full-field visual stimuli were presented on a video screen (SONY, Pencoed, UK) in 120 Hz noninterlaced mode. High contrast, square-wave gratings were generated at optimal spatial (0.1 -- 0.2 cycle/deg) and temporal frequencies (1 -- 2 Hz), using the stimulus generator systems VSG Series Three (Cambridge Research Systems, Rochester, UK). After the recording session, the animals were euthanized with an overdose of anesthetics (Nembuthal). High vs. low spatial frequency stimuli characteristic, respectively, for area 17 and area 18 were used to visualize the area border.\\
Difference maps were obtained from single-condition maps as described previously \cite{Buzas1998, Yousef1999}.  For each hemisphere a region of interest (ROI) was defined containing the imaged part of area 17. ROIs typically contained around 10-20$\Lambda^2$ were $\Lambda$ is the typical column spacing of the OPM. Each raw difference map was Fermi-bandpass filtered as in \cite{Kaschube2010, Keil2012}. Low-pass cut-off was chosen as $1.5\text{mm}$, high-pass cut-off was chosen as $0.4\text{mm}$. This preprocessing ensured efficient removal of high-frequency noise from the CCD-camera and low frequency variations in signal strength while only weakly attenuating structures on the relevant scales. 
\subsection*{Experimental mosaics}
All RGC mosaics used in the present study are available for download from the website of Dr. Stephen J. Eglen (http://www.damtp.cam.ac.uk/user/sje30/data/mosaics/).
\subsection*{Numerical procedures for mPIPP mosaics}
%%%
%
%
%
%%%%%%%%%%%%%%%%% FIGURE %%%%%%%%%%%%%%%
%
%
%
\begin{figure}
\begin{center}
\includegraphics[width=12cm]{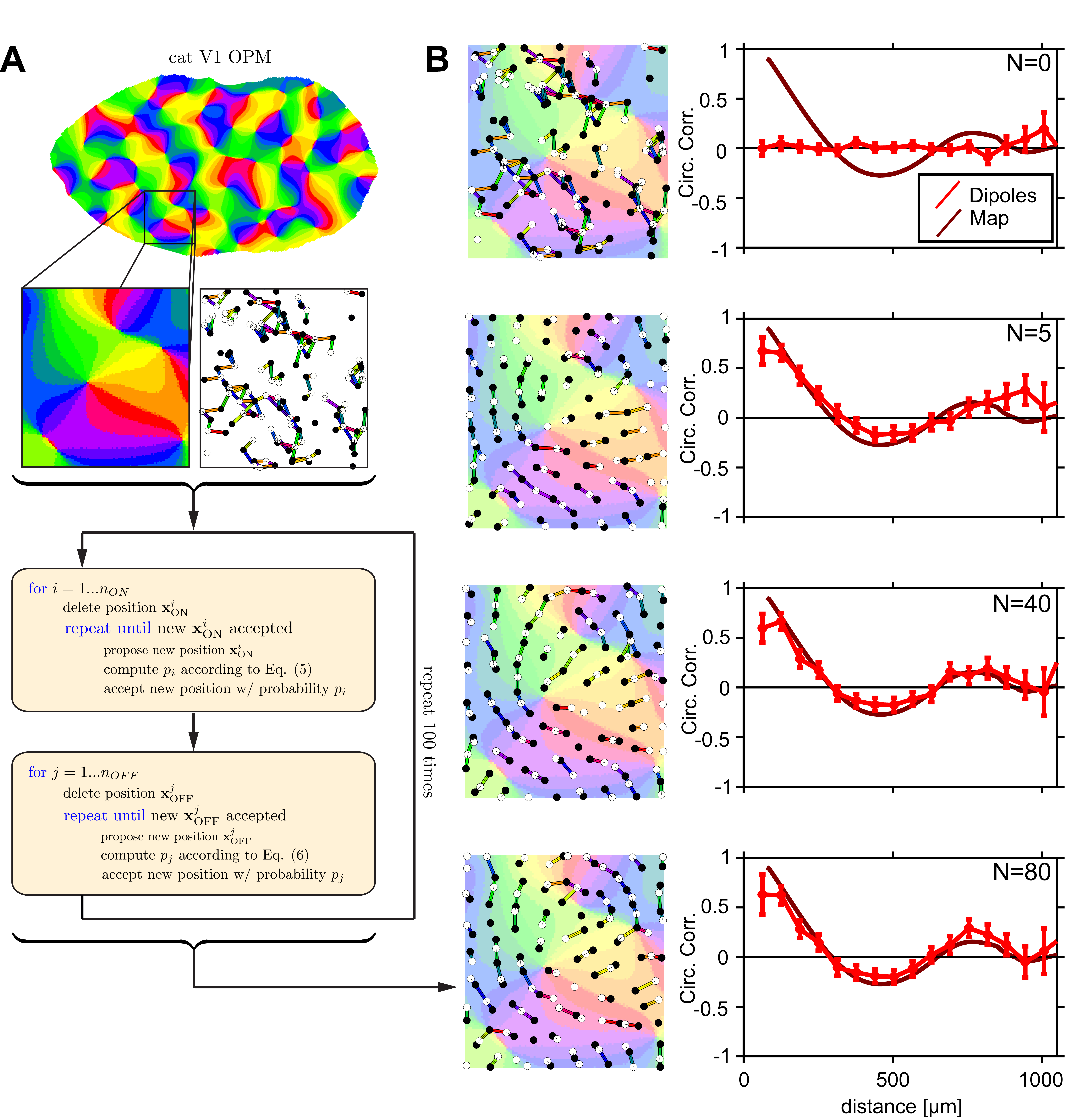}
\end{center}
\caption{\textbf{Constructing RGC mosaics from measured OPMs with an mPIPP}. Upper panel: OPM measured in cat V1. Below, left: Inset of OPM used for modulating the mPIPP; right: Initial condition of the Monte-Carlo optimization procedure. ON/OFF RGC mosaic are specified by a homogeneous Poisson process. Positioning of RGCs is irregular. The preferred orientations of dipoles (colored bars) do not match the orientation preferences of the OPM region. Lower most panel:  ON/OFF RGC mosaic after 20 iterations of the Monte-Carlo procedure with $\gamma = 30$ and $\xi=0.6\text{mm}_c/\text{mm}_r$ for mosaic m623 (see Materials and Methods). Positioning of RGCs is semiregular and preferred orientation of dipoles are almost perfectly aligned with orientation preferences of the OPM region (background). Colorcode as in Fig. \ref{fig1}C.
\textbf{B} Orientation map region, superimposed mPIPP mosaics (left) and corresponding dipole correlation function (right)  for 0,5,40, and 80 iterations (upper most to lower most panels). Note that the dipole angle correlations settle around their final values within only a few iterations of the process, whereas the precise positions of the ON/OFF cells continues to change. 
\label{mPIPP_simulations}}
\end{figure}
%%%%%%%%%%%%%%%%%%%%%%%%%%%%%%%%%%
%
%
Numerical procedures to obtain mPIPP ON/OFF mosaics were chosen as previously described \cite{Eglen2005, Ringach2007}. In short, we initially positioned $n_\text{OFF}$ OFF-cells and $n_\text{ON}$ ON-cells independently according to a two-dimensional Poisson point process with cell density matched to the density of cells in m623 and w81s1 respectively. We then updated these positions according to the following loop (see Fig. \ref{mPIPP_simulations}A):
For each ON-center cell a new candidate position was generated at random. Considering the i-th ON-center cell, this new position was accepted with probability
\begin{eqnarray}
p_i=\prod_{j=1,i\neq j}^{n_\text{ON}} h_{\text{ON},\text{ON}}(|\mathbf{x}^i_\text{ON}-\mathbf{x}^j_\text{ON}|)\cdot \prod_{j=1}^{n_\text{OFF}}H_{\text{ON},\text{OFF}}(\mathbf{x}^i_\text{ON},\mathbf{x}^j_\text{OFF})\,,
\label{eq:p_i_ON}
\end{eqnarray}
where $H_{\text{ON},\text{OFF}}$ is defined as in Eq. \eqref{eq:intra-interaction_function_definition}.
After updating all ON center cells' positions, the procedure was repeated for the OFF center cells, here for the cell number $i$
\begin{eqnarray}
p_i=\prod_{j=1,i\neq j}^{n_\text{OFF}} h_{\text{OFF}, \text{OFF}}(|\mathbf{x}^i_\text{OFF}-\mathbf{x}^j_\text{OFF}|) \cdot\prod_{j=1}^{n_\text{ON}}H_{\text{ON},\text{OFF}}(\mathbf{x}^j_\text{ON},\mathbf{x}^i_\text{OFF})
\label{eq:p_i_OFF}
\end{eqnarray}
with different parameters for $h_{\text{OFF},\text{OFF}}$ and $h_{\text{ON},\text{ON}}$. Both loops were repeated between 20 and 50 times. Note that from one complete iteration of the algorithm to the next, the absolute positions of all cells show no correlation (cf. Fig. \ref{mPIPP_simulations}). However, the spatial statistics of the pattern including dipole angle correlations are stable characteristics of the mosaics after only a few iterations.

Model parameters besides $\gamma$ and $d$ were chosen as to match the local spatial statistics of m623, w81s1 \cite{Eglen2005}  and g09 \cite{Hore2012}, respectively, and are summarized in Supplementary Table 1. RGCs close to the boundary of the simulated domain can only form dipoles in a subset of all possible directions, potentially leading to boundary effects in the mPIPP simulations. Since the dipole orientation is enforced by a given OPM, we usually observe a slight decrease in the dipole density towards the very edge of the retinal patches. However, for the rather small beta-values that are consistent with the experimental data (cf. Fig. \ref{fig5}), such boundary effects were observed to play a negligible role.

\subsection*{Dipole extraction and correlation function}
Following \cite{Paik2012b}, we assumed a pair of ON/OFF ganglion cells at position $\mathbf{x}$ and $\mathbf{y}$ respectively to form a dipole if their distance was smaller than a parameter $d$, i.e. $ \|\mathbf{x}-\mathbf{y} \|_2 < d $. A dipole's preferred orientation was defined as in Eq. (\ref{Eq:dipole_def}). Note that this orientation is orthogonal to the orientation of the dipole vector connecting the ON/OFF pair (see Fig. \ref{fig1}A). A dipole's position was defined as $(\mathbf{x}+\mathbf{y})/2$, i.e. half way between the ON and the OFF cell. Note that one RGC can form multiple dipoles depending on the choice of $d$. 
Spatial correlations $C(R)$ between dipole angles $\phi(\mathbf{x})$ and $\phi(\mathbf{y})$ were calculated as \cite{Wolf2003,Schnabel2007}
$$
C(R) = \left< \cos \left(\phi(\mathbf{x}) - \phi(\mathbf{y})\right) \right>_{(R  - b/2) \leq  \|\mathbf{x} - \mathbf{y} \|_2  <( R +b/2)}\,,
$$
where $b$ is a fixed bin size. To obtain Fig. \ref{fig2}B, Figs. \ref{fig4}G-J and Fig. \ref{fig5}, for a given mosaic, the bin size was chosen such that the diagonal of the rectangular retinal section considered contained $20$ equidistant bins.
Cross-correlation between the modulating OPM $\theta(\mathbf{x})$ and the dipole orientations $\phi(\mathbf{x})$ (Fig. \ref{fig1}D, lower panel) were determined with
$$
C_\text{cross} = \left<\cos\left(\phi(\mathbf{x}) - \theta(\mathbf{x})\right)\right>_{\text{dipoles}}\,.
$$
where the average is taken over all dipoles in the retinal region considered.
The expected value of $C_\text{cross}$ for an mPIPP with a given $\gamma$ can be estimated using the probability $p$ of positioning an ON/OFF cell 
$$
p(\Delta \theta)=\frac{1}{N}\exp(\gamma( \cos(\Delta \theta)-1) )\,,
$$ where $\Delta \theta$ is the difference between the dipole orientation $\phi$ and the preferred orientation specified in the orientation map $\theta (\mathbf{x})$ and $N$ is a normalizing factor, i.e. $N=\int d(\Delta \theta) \exp(\gamma(\cos(\Delta \theta)-1) )$. For large enough retinal regions, we have 
$$
\left<\cos(\Delta \theta)\right>_{\text{dipoles}} =  \int_{-\pi}^{\pi} d(\Delta \theta) p(\Delta \theta) \cos(\Delta \theta) = \frac{I_2(\gamma)}{I_0(\gamma)}\,,
$$
where $I_j(\gamma)$ is the j-th modified Bessel function of the first kind. This function is depicted in Fig. \ref{fig2}C (blue line) along with the numerically obtained cross-correlation values (red line). 
\newline
\subsection*{Cortical magnification factor at the position of the measured mosaics.}
To determine the V1 OPM region sizes to which the two published cat beta cell mosaics correspond, we estimated the cortical magnification factor at the respective retinal positions. The center of mosaic w81s1 was located $19^\circ$ below the mid line of the visual streak, $4mm$ from the area centralis \cite{Zhan2000}. With respect to a cartesian coordinate system with origin at the area centralis and x-axis along the visual streak this position is 
\begin{eqnarray*}
 \begin{pmatrix} 4\text{ mm}_r\cos(19^\circ) \\ 4\text{ mm}_r\sin(19^\circ) \end{pmatrix}=\begin{pmatrix} 1.3\text{ mm}_r \\ 3.8\text{ mm}_r \end{pmatrix}
\end{eqnarray*}
In the cat retina, $1\text{ mm}_r$ roughly corresponds to $4.4\text{ deg}$ visual angle \cite{Barlow1957, Bishop1962}. Thus
\begin{eqnarray*}
\begin{pmatrix} 1.3\text{ mm}_r \\ 3.8\text{ mm}_r \end{pmatrix}\to \begin{pmatrix} 5.9^\circ \\ 17.2^\circ \end{pmatrix}.
\end{eqnarray*}
From \cite{Tusa1978}, we determined the cortical magnification for elevation $5.9^\circ$ and azimuth $17.2^\circ$ to be about $0.15\text{mm}_c^2/\text{deg}^2$. Thus, via 
\begin{eqnarray*}
 \xi=\sqrt{0.15\frac{\text{mm}_c^2}{\text{deg}^2}}\cdot4.4\frac{\text{deg}}{\text{mm}_r}=1.7\frac{\text{mm}_c}{\text{mm}_r}\,,
\end{eqnarray*}
we estimate that, at the location of w81s1, $0.6mm$ on the retina correspond to $1mm$ visual cortex. The typical column spacing of cat OPMs is $\Lambda_c\approx 1\text{ mm}_c$ \cite{Rathjen2003, Kaschube2002}. Hence, we would expect a periodicity of about $\Lambda_r=0.6\text{mm}_r$ on the retina. \newline
The mosaic m623 has been obtained from 5mm eccentricity and 5.5 deg below the mid line of the visual streak. This corresponds to the point $(2.1^\circ, 22.6^\circ)$ in the visual field, which again has $0.15\text{mm}_c^2/\text{deg}^2$ cortical magnification.
Thus $0.6\text{mm}_r \, \widehat{=} \, 1\text{mm}_c$ for m623.\newline
The mosaic g09 was measured at 9mm eccentricity (temporal retina) at 41$^\circ$ visual angle in Macaca mulatta, Macaca fascicularis. At this point $1.4 \text{mm}_r  \widehat{=}  1\text{mm}_c$ (\cite{Paik2011}, Suppl. Inf.).

\subsection*{Spatial statistics of mPIPP mosaics}
We have analyzed the spatial statistics of mPIPP mosaics using previous methods \cite{Eglen2005,Hore2012}. In short, to obtain the $G$-function, for each point in a mosaic the distance to its nearest neighbor was calculated and the cumulative distribution of these distances was computed. The $L$-function is the scaled expectation of the number of points observed within a given distance of any point. In Figures \ref{eglen_stat_analysis_off}C and D,  we have drawn the 95\% confidence levels from 99 mPIPP simulations (gray shading). Informally, if the measure from the observed mosaics (solid lines in Fig. \ref{eglen_stat_analysis_off}) falls within the confidence intervals, then the model is a good fit to the observed mosaic. Topological disorder, $\mu_2$ was quantified using \cite{Kram2010}
$$
\mu_2 = \sum_n (n-6)^2P_n\,,
$$
where $P_n$ is the probability of a Voronoi polygon having $n$ edges.
\subsection*{Calculation of p-values for local spatial statistics}
All p-values used in the present study are Monte Carlo P values  \cite{Diggle1986, Eglen2005}. 
Each model is run, using the same parameters but with different initial conditions, 99 times. Each mosaic (both observed and simulated) is then compared against the other 99 mosaics, and a certain measure T is calculated. The 99 T values are sorted, and the rank of the T measure corresponding to the observed field is divided by 99 to calculate a p-value. p-values of 0.05 and smaller indicate that the model does not fit the observed data at the 5\% significance level. The smallest p-value calculated is thus 0.01 with 99 simulations.
\subsection*{Calculation of p-values for spatial correlation of dipole angles}
From $N = 100$ independent realizations of the mPIPP, we obtained circular correlation functions $C^i(R)$ with $i \in [1,100]$ and $R$ being the binned distance (see above). We introduced a measure 
\begin{equation}
T_i = \sum_{R = 0 \mu m}^{200 \mu m} \left(C^i(R) - \left<C(R)\right>\right)^2
\label{eq:T_definition}
\end{equation}
where $\left<C(R)\right> = \frac{1}{N} \sum_i C^i(R) $ is is the average correlation of the mPIPP simulations. Note that only bins with $R \leq 200\mu m$ were used (the region of strong positive correlation). The distribution of values T quantifies how much the test ensemble of $C_i$ deviates from the average. Next, we repeated this evaluation with the correlation function $M(R)$ obtained from the cat mosaics:
\begin{equation}
S = \sum_{R = 0 \mu m}^{200 \mu m} \left(M(R) - \left<C(R)\right>\right)^2
\label{eq:S_definition}
\end{equation}
From the distribution of  the $T_i$'s and value of S, the Monte-Carlo p-value was estimated as described above. 
\subsection*{Estimation of the statistical power of hypothesis test}
The null hypothesis of the statistical test is that RGC dipole angles are correlated in space. To estimate the statistical power of our hypothesis test for the presence of correlations, it is necessary to estimate the probability that the test will reject the presence of correlations when the alternative hypothesis is true, i.e. the RGC mosaics stems from an ensemble of correlations. To do so, we simulated an ensemble of mPIPP mosaics with known correlational statistics. Different realizations of the mPIPP were then compared to a PIPP control ensemble with respect to their dipole angle correlations. In some mPIPP realizations, the angular correlation of the mPIPP will be within the range of correlations of the PIPP ensemble. Comparing these correlations values and concluding that the mPIPP realization stems from a PIPP ensemble without angular correlations would constitute a type II statistical error (false negative). The probability of such an error, i.e. false negative rate, is a measure of the statistical power of the test. This probability was estimated by repeatedly comparing mPIPP realizations to PIPP ensembles.\newline
More precisely, we first chose a value $\gamma$, an area size $A$ and a number of mosaics $N$. Within the area $A$ of the retinal patch, the RGC density was fixed to the values of mosaic m623, $\rho_{\text{ON}} = 67\text{ mm}^{-2} $ and $\rho_{\text{OFF}} = 74\text{ mm}^{-2} $. Then, we calculated 500 realizations of mosaics for this area and $\gamma=0$ to obtain the PIPP ensemble. From this ensemble, we randomly drew 100 mosaics with replacement and calculated their correlation functions. From the 100 correlation functions, N were drawn and averaged, in total a 100 times. From this set of averaged correlation functions we calculate the distribution of $T_i$ according to Eq. (\ref{eq:T_definition}). Next, we compare these $T_i$ with an S (Eq. (\ref{eq:S_definition})) obtain from averaging correlations functions of N random mPIPP realizations of the chosen $\gamma$ and with the same $A$. This resulted in a single p-value for this particular set of realizations.  To estimate the probability of falsely rejecting the presence of angular dipole correlation, we calculated a distribution of such p-values by repeating the above stated steps a 1000 times. p-values larger than 0.05 (our significance level) in this distribution indicate the occurrence of type II statistical error. Fig. \ref{fig:error_II_test}A shows the percentage of p-values larger than 0.05 (i.e. the false negative rate, $\beta$) as function of $A$ and $N$ for different values of $\gamma$. Fig. \ref{fig:error_II_test}A  shows the same analysis for periodicity. It is done analogously with correlations evaluated between 200$\mu m$ and 400$\mu m$.
\clearpage

% Do NOT remove this, even if you are not including acknowledgments
\section*{Acknowledgments}
We are grateful to Z. Kisvarday (University of Debrecen, Debrecen, Hungary) for sharing optical imaging data. We thank Ana Ho\v{c}evar Brezav\v{s}\v{c}ek (Center for Studies in Physics and Biology, The Rockefeller University, New York, NY, USA) for fruitful discussions and detailed comments on an earlier version of this manuscript.
%
%\section*{References}
% The bibtex filename
\bibliography{./library_Schottdorf_et_al}

\begin{thebibliography}{10}
\providecommand{\url}[1]{\texttt{#1}}
\providecommand{\urlprefix}{URL }
\expandafter\ifx\csname urlstyle\endcsname\relax
  \providecommand{\doi}[1]{doi:\discretionary{}{}{}#1}\else
  \providecommand{\doi}{doi:\discretionary{}{}{}\begingroup
  \urlstyle{rm}\Url}\fi
\providecommand{\bibAnnoteFile}[1]{%
  \IfFileExists{#1}{\begin{quotation}\noindent\textsc{Key:} #1\\
  \textsc{Annotation:}\ \input{#1}\end{quotation}}{}}
\providecommand{\bibAnnote}[2]{%
  \begin{quotation}\noindent\textsc{Key:} #1\\
  \textsc{Annotation:}\ #2\end{quotation}}
\providecommand{\eprint}[2][]{\url{#2}}

\bibitem{Paik2011}
Paik SB, Ringach DL (2011) {Retinal origin of orientation maps in visual
  cortex.}
\newblock Nature Neuroscience 14: 919--925.
\bibAnnoteFile{Paik2011}

\bibitem{Hubel1962}
Hubel DH, Wiesel TN (1962) {Receptive fields, binocular interaction and
  functional architecture in the cat's visual cortex.}
\newblock The Journal of Physiology 160: 106--154.
\bibAnnoteFile{Hubel1962}

\bibitem{Ohki2006}
Ohki K, Chung S, Kara P, H\"{u}bener M, Bonhoeffer T, et~al. (2006) {Highly
  ordered arrangement of single neurons in orientation pinwheels.}
\newblock Nature 442: 925--928.
\bibAnnoteFile{Ohki2006}

\bibitem{Blasdel1992a}
Blasdel GG (1992) {Orientation selectivity, preference, and continuity in
  monkey striate cortex.}
\newblock The Journal of Neuroscience 12: 3139--3161.
\bibAnnoteFile{Blasdel1992a}

\bibitem{Chapman1996}
Chapman B, Stryker MP, Bonhoeffer T (1996) {Development of orientation
  preference maps in ferret primary visual cortex.}
\newblock The Journal of Neuroscience 16: 6443--6453.
\bibAnnoteFile{Chapman1996}

\bibitem{Crair1997}
Crair MC, Ruthazer ES, Gillespie DC, Stryker MP (1997) {Ocular dominance peaks
  at pinwheel center singularities of the orientation map in cat visual
  cortex.}
\newblock The Journal of Neurophysiology 77: 3381--3385.
\bibAnnoteFile{Crair1997}

\bibitem{Obermayer1997}
Obermayer K, Blasdel GG (1997) {Singularities in primate orientation maps.}
\newblock Neural Computation 9: 555--575.
\bibAnnoteFile{Obermayer1997}

\bibitem{Bosking1997}
Bosking WH, Zhang Y, Schofield B, Fitzpatrick D (1997) {Orientation selectivity
  and the arrangement of horizontal connections in tree shrew striate cortex.}
\newblock The Journal of Neuroscience 17: 2112--2127.
\bibAnnoteFile{Bosking1997}

\bibitem{Shmuel2000}
Shmuel A, Grinvald A (2000) {Coexistence of linear zones and pinwheels within
  orientation maps in cat visual cortex.}
\newblock Proceedings of the National Academy of Sciences, USA 97: 5568--5573.
\bibAnnoteFile{Shmuel2000}

\bibitem{Kaschube2009}
Kaschube M, Schnabel M, Wolf F, L\"{o}wel S (2009) {Interareal coordination of
  columnar architectures during visual cortical development.}
\newblock Proceedings of the National Academy of Sciences, USA 106:
  17205--17210.
\bibAnnoteFile{Kaschube2009}

\bibitem{Kaschube2010}
Kaschube M, Schnabel M, L\"{o}wel S, Coppola DM, White LE, et~al. (2010)
  {Universality in the evolution of orientation columns in the visual cortex.}
\newblock Science 330: 1113--1116.
\bibAnnoteFile{Kaschube2010}

\bibitem{Keil2012}
Keil W, Kaschube M, Schnabel M, Kisvarday ZF, L{\"o}wel S, et~al. (2012)
  {Response to Comment on "Universality in the Evolution of Orientation Columns
  in the Visual Cortex"}.
\newblock Science 336: 413--413.
\bibAnnoteFile{Keil2012}

\bibitem{Durbin1990}
Durbin R, Mitchison G (1990) {A dimension reduction framework for understanding
  cortical maps.}
\newblock Nature 343: 644--647.
\bibAnnoteFile{Durbin1990}

\bibitem{Obermayer1992}
Obermayer K, Blasdel GG, Schulten K (1992) {Statistical-mechanical analysis of
  self-organization and pattern formation during the development of visual
  maps.}
\newblock Physical Review A 45: 7568--7589.
\bibAnnoteFile{Obermayer1992}

\bibitem{Swindale1996}
Swindale NV (1996) {The development of topography in the visual cortex: a
  review of models.}
\newblock Network 7: 161--247.
\bibAnnoteFile{Swindale1996}

\bibitem{Wolf1998}
Wolf F, Geisel T (1998) {Spontaneous pinwheel annihilation during visual
  development.}
\newblock Nature 395: 73--78.
\bibAnnoteFile{Wolf1998}

\bibitem{Wolf2005}
Wolf F (2005) {Symmetry, Multistability, and Long-Range Interactions in Brain
  Development.}
\newblock Physical Review Letters 95: 208701.
\bibAnnoteFile{Wolf2005}

\bibitem{Keil2011}
Keil W, Wolf F (2011) {Coverage, continuity, and visual cortical architecture.}
\newblock Neural Systems \& Circuits 1: 1--17.
\bibAnnoteFile{Keil2011}

\bibitem{Ringach2004a}
Ringach DL (2004) {Haphazard wiring of simple receptive fields and orientation
  columns in visual cortex.}
\newblock The Journal of Neurophysiology 92: 468--476.
\bibAnnoteFile{Ringach2004a}

\bibitem{Ringach2007}
Ringach DL (2007) {On the origin of the functional architecture of the cortex.}
\newblock PLoS One 2: e251.
\bibAnnoteFile{Ringach2007}

\bibitem{Paik2012}
Paik SB, Ringach DL (2012) {Link between orientation and retinotopic maps in
  primary visual cortex.}
\newblock Proceedings of the National Academy of Sciences, USA 109: 7091--7096.
\bibAnnoteFile{Paik2012}

\bibitem{Soodak1986}
Soodak RE (1986) {Two-dimensional modeling of visual receptive fields using
  Gaussian subunits.}
\newblock Proceedings of the National Academy of Sciences, USA 83: 9259--9263.
\bibAnnoteFile{Soodak1986}

\bibitem{Soodak1987}
Soodak RE (1987) {The retinal ganglion cell mosaic defines orientation columns
  in striate cortex.}
\newblock Proceedings of the National Academy of Sciences, USA 84: 3936--3940.
\bibAnnoteFile{Soodak1987}

\bibitem{Cleland1971}
Cleland BG, Dubin MW, Levick WR (1971) {Simultaneous recording of input and
  output of lateral geniculate neurones.}
\newblock Nature New Biology 231: 191--192.
\bibAnnoteFile{Cleland1971}

\bibitem{Cleland1985}
Cleland BG, Lee BB (1985) {A comparison of visual responses of cat lateral
  geniculate nucleus neurones with those of ganglion cells afferent to them.}
\newblock The Journal of Physiology 369: 249--268.
\bibAnnoteFile{Cleland1985}

\bibitem{Daniel1961a}
Daniel PM, Whitteridge D (1961) {The representation of the visual field on the
  cerebral cortex in monkeys.}
\newblock The Journal of Physiology 159: 203--221.
\bibAnnoteFile{Daniel1961a}

\bibitem{Tusa1978}
Tusa RJ, Palmer LA, Rosenquist AC (1978) {The retinotopic organization of area
  17 (striate cortex) in the cat.}
\newblock The Journal of Comparative Neurology 177: 213--235.
\bibAnnoteFile{Tusa1978}

\bibitem{Tootell1982}
Tootell RB, Silverman MS, Switkes E, Valois RL (1982) {Deoxyglucose analysis of
  retinotopic organization in primate striate cortex.}
\newblock Science 218: 902--904.
\bibAnnoteFile{Tootell1982}

\bibitem{Djavadian1983}
Djavadian RL, Harutiunian-Kozak BA (1983) {Retinotopic organization of the
  lateral suprasylvian area of the cat.}
\newblock Acta Neurobiologiae Experimentalis 43: 251--262.
\bibAnnoteFile{Djavadian1983}

\bibitem{Wassle1981a}
W\"{a}ssle H, Boycott BB, Illing RB (1981) {Morphology and mosaic of on- and
  off-beta cells in the cat retina and some functional considerations.}
\newblock Proceedings of the Royal Society of London Series B 212: 177--195.
\bibAnnoteFile{Wassle1981a}

\bibitem{Alonso2001}
Alonso JM, Usrey WM, Reid RC (2001) {Rules of connectivity between geniculate
  cells and simple cells in cat primary visual cortex.}
\newblock The Journal of Neuroscience 21: 4002--4015.
\bibAnnoteFile{Alonso2001}

\bibitem{Amidror2009}
Amidror I (2009) {The Theory of the Moir\'{e} Phenomenon, Volume I}.
\newblock Springer Verlag.
\bibAnnoteFile{Amidror2009}

\bibitem{Hore2012}
Hore VRA, Troy JB, Eglen SJ (2012) {Parasol cell mosaics are unlikely to drive
  the formation of structured orientation maps in primary visual cortex.}
\newblock Visual Neuroscience 29: 283-299.
\bibAnnoteFile{Hore2012}

\bibitem{Eglen2005}
Eglen SJ, Diggle PJ, Troy JB (2005) {Homotypic constraints dominate positioning
  of on-and off-center beta retinal ganglion cells.}
\newblock Visual Neuroscience 22: 859--871.
\bibAnnoteFile{Eglen2005}

\bibitem{Zhan2000}
Zhan XJ, Troy JB (2000) {Modeling cat retinal beta-cell arrays.}
\newblock Visual Neuroscience 17: 23--39.
\bibAnnoteFile{Zhan2000}

\bibitem{Gauthier2009}
Gauthier JL, Field GD, Sher A, Greschner M, Shlens J, et~al. (2009) {Receptive
  fields in primate retina are coordinated to sample visual space more
  uniformly.}
\newblock PLoS Biology 7: e1000063.
\bibAnnoteFile{Gauthier2009}

\bibitem{Paik2012b}
Paik SB, Li PH, Chichilnisky EJ, Ringach DL (2012) {Analysis of ON/OFF-dipole
  spatial statistics in retinal ganglion cell mosaics.}
\newblock In: Society for Neuroscience Annual Meeting.
\bibAnnoteFile{Paik2012b}

\bibitem{Bonhoeffer1991}
Bonhoeffer T, Grinvald A (1991) {Iso-orientation domains in cat visual cortex
  are arranged in pinwheel-like patterns.}
\newblock Nature 353: 429--431.
\bibAnnoteFile{Bonhoeffer1991}

\bibitem{Kaschube2002}
Kaschube M, Wolf F, Geisel T, L\"{o}wel S (2002) {Genetic influence on
  quantitative features of neocortical architecture.}
\newblock The Journal of Neuroscience 22: 7206--7217.
\bibAnnoteFile{Kaschube2002}

\bibitem{Kram2010}
Kram YA, Mantey S, Corbo JC (2010) {Avian cone photoreceptors tile the retina
  as five independent, self-organizing mosaics.}
\newblock PLoS One 5: e8992.
\bibAnnoteFile{Kram2010}

\bibitem{Usrey1999}
Usrey WM, Reppas JB, Reid RC (1999) {Specificity and strength of
  retinogeniculate connections.}
\newblock The Journal of Neurophysiology 82: 3527--3540.
\bibAnnoteFile{Usrey1999}

\bibitem{Crair1998}
Crair MC, Gillespie DC, Stryker MP (1998) {The Role of Visual Experience in the
  Development of Columns in Cat Visual Cortex.}
\newblock Science 279: 566--570.
\bibAnnoteFile{Crair1998}

\bibitem{Hensch2004}
Hensch TK, Stryker MP (2004) Columnar architecture sculpted by \textsc{GABA}
  circuits in developing cat visual cortex.
\newblock Science 303: 1678-1681.
\bibAnnoteFile{Hensch2004}

\bibitem{vonderMalsburg1973}
von~der Malsburg C (1973) {Self-organization of orientation sensitive cells in
  the striate cortex}.
\newblock Kybernetik 14: 85--100.
\bibAnnoteFile{vonderMalsburg1973}

\bibitem{Swindale1982a}
Swindale NV (1982) {A model for the formation of orientation columns.}
\newblock Proceedings of the Royal Society of London Series B 215: 211--230.
\bibAnnoteFile{Swindale1982a}

\bibitem{Keil2010}
Keil W, Schmidt KF, L{\"o}wel S, Kaschube M (2010) {Reorganization of columnar
  architecture in the growing visual cortex}.
\newblock Proceedings of the National Academy of Sciences, USA 107:
  12293--12298.
\bibAnnoteFile{Keil2010}

\bibitem{Adams2002}
Adams DL, Horton JC (2002) {Shadows cast by retinal blood vessels mapped in
  primary visual cortex.}
\newblock Science 298: 572--576.
\bibAnnoteFile{Adams2002}

\bibitem{Giacomantonio2007}
Giacomantonio CE, Goodhill GJ (2007) {The effect of angioscotomas on map
  structure in primary visual cortex.}
\newblock J Neurosci 27: 4935--4946.
\bibAnnoteFile{Giacomantonio2007}

\bibitem{Buzas1998}
Buz\'{a}s P, Eysel UT, Kisv\'{a}rday ZF (1998) {Functional topography of single
  cortical cells: An intracellular approach combined with optical imaging.}
\newblock Brain Research Protocols 3: 199--208.
\bibAnnoteFile{Buzas1998}

\bibitem{Yousef1999}
Yousef T, Bonhoeffer T, Kim DS, Eysel UT, T\'{o}th E, et~al. (1999)
  {Orientation topography of layer 4 lateral networks revealed by optical
  imaging in cat visual cortex (area 18).}
\newblock European Journal of Neuroscience 11: 4291--4308.
\bibAnnoteFile{Yousef1999}

\bibitem{Wolf2003}
Wolf F, Geisel T (2003) {Universality in visual cortical pattern formation.}
\newblock The Journal of Physiology 97: 253--264.
\bibAnnoteFile{Wolf2003}

\bibitem{Schnabel2007}
Schnabel M, Kaschube M, L\"{o}wel S, Wolf F (2007) {Random waves in the brain:
  Symmetries and defect generation in the visual cortex.}
\newblock The European Physical Journal Special Topics 145: 137--157.
\bibAnnoteFile{Schnabel2007}

\bibitem{Barlow1957}
Barlow HB, Fitzhugh R, Kuffler SW (1957) {Change of organization in the
  receptive fields of the cat's retina during dark adaptation.}
\newblock The Journal of Physiology 137: 338--354.
\bibAnnoteFile{Barlow1957}

\bibitem{Bishop1962}
Bishop PO, Kozak W, Vakkur GJ (1962) {Some quantitative aspects of the cat's
  eye: axis and plane of reference, visual field co-ordinates and optics.}
\newblock The Journal of Physiology 163: 466--502.
\bibAnnoteFile{Bishop1962}

\bibitem{Rathjen2003}
Rathjen S, Schmidt KE, L\"{o}wel S (2003) {Postnatal growth and column spacing
  in cat primary visual cortex.}
\newblock Experimental Brain Research 149: 151--158.
\bibAnnoteFile{Rathjen2003}

\bibitem{Diggle1986}
Diggle PJ (1986) {Displaced amacrine cells in the retina of a rabbit: analysis
  of a bivariate spatial point pattern.}
\newblock Journal of Neuroscience Methods 18: 115--125.
\bibAnnoteFile{Diggle1986}

\end{thebibliography}
\section*{Supplementary Table 1}
\begin{tabular}{|r||r|r|r|r|r|r|r|r|r|} \hline
\textbf{Mosaic key} & \multicolumn{9}{c|}{\textbf{Parameters}}\\
\hline
 & $n_{\text{ON}}$ & $n_{\text{OFF}}$ & $l_x [\mu m]$ & $l_y [\mu m]$ & $\varphi_{\text{ON}}  [\mu m]$ & $\varphi_{\text{OFF}}  [\mu m]$ &$\alpha_{\text{ON}}$ & $\alpha_{\text{OFF}}$ & $\delta [\mu m]$  \\
\cline{1-10}
w81S1 & 65 & 70  & 750 & 991 & 67.94 & 66.27 & 7.81 & 5.40 & 18 \\ \hline
m623 & 74 & 82  & 1002 &  1100 & 112.79 & 65.46 & 3.05 & 8.11 & 20 \\ \hline
G09 & 89 & 117  & 1850 &  1075 & 130 & 125 & 14.5 & 13.0 & 20 \\ \hline
\end{tabular}
\clearpage
\section*{Supplementary Figures}
\begin{figure}
\begin{center}
\includegraphics[width=12cm]{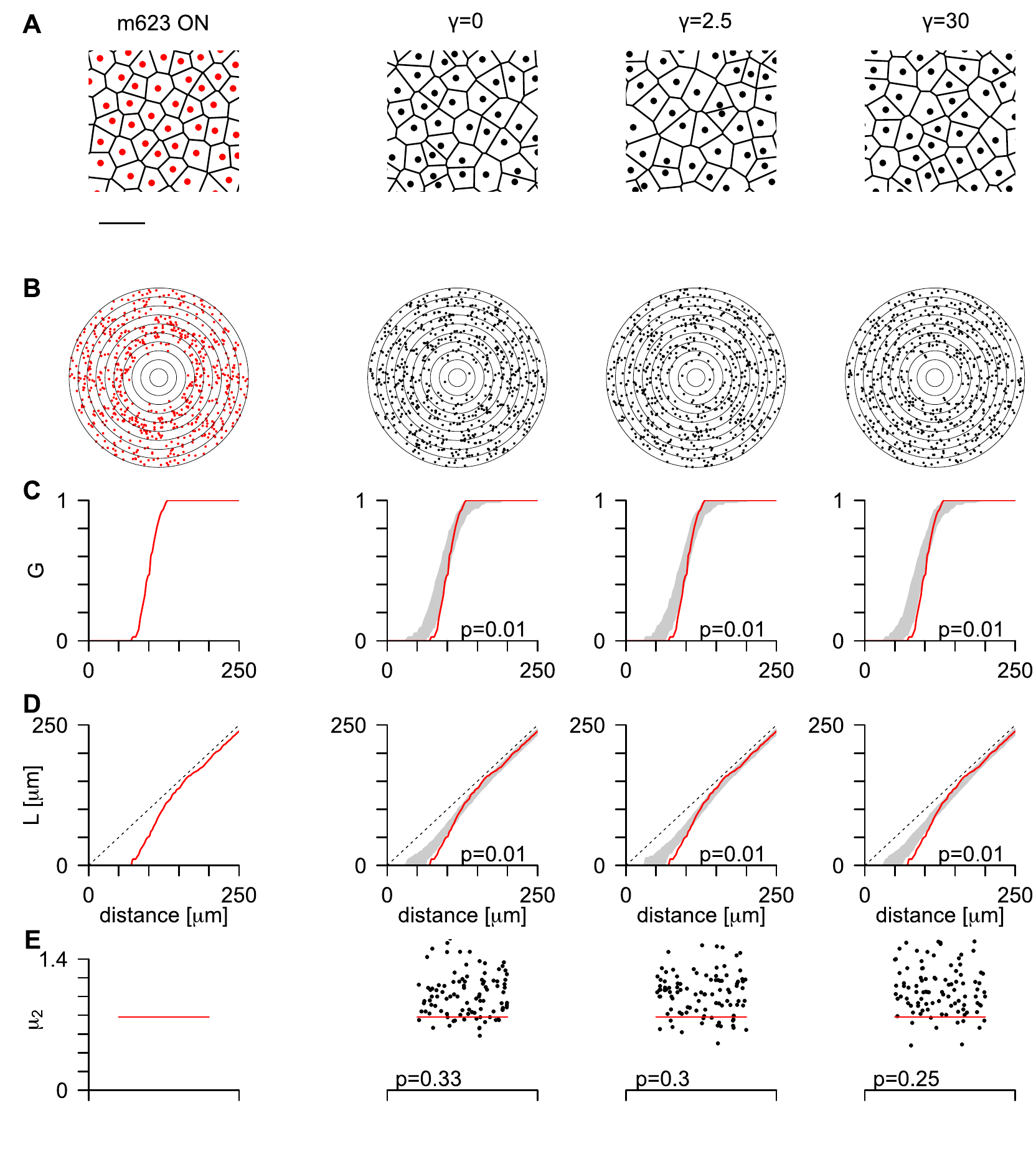}
\end{center}
\caption{\textbf{Spatial properties of ON cell positions in mPIPP mosaics are independent of $\gamma$ and in agreement with experimental data.}
Same as Fig. \ref{eglen_stat_analysis_off}, but for ON cells in RGC mosaic m623.
\label{ON_cell_spatial_statistics}}
\end{figure}
%%
%
%
%
%
%%%%%%%%%%%%%%%%% FIGURE %%%%%%%%%%%%%%%%%

%
%
%\begin{figure}[!ht]
%\begin{center}
%%%\includegraphics[width=4in]{figure_name.2.eps}
%\end{center}
%\caption{
%{\bf Bold the first sentence.}  Rest of figure 2  caption.  Caption 
%should be left justified, as specified by the options to the caption 
%package.
%}
%\label{Figure_label}
%\end{figure}
%\section*{Tables}
%\begin{table}[!ht]
%\caption{
%\bf{Table title}}
%\begin{tabular}{|c|c|c|}
%table information
%\end{tabular}
%\begin{flushleft}Table caption
%\end{flushleft}
%\label{tab:label}
% \end{table}

\end{document}